\documentclass[manuscript,screen]{acmart}
\usepackage{graphicx}
\usepackage{textcomp}
\usepackage{multirow}
\usepackage{booktabs}
\usepackage{tabularx}
\usepackage{booktabs}

\AtBeginDocument{%
  }

\setcopyright{acmcopyright}
\copyrightyear{}
\acmYear{}
\acmDOI{XXXXXXX.XXXXXXX}

\acmPrice{15.00}
\acmISBN{978-1-4503-XXXX-X/18/06}

\begin{document}

\title{Disease Outbreak Detection and Forecasting: A  Review of Methods and Data Sources}

\author{Ghazaleh Babanejaddehaki}
\email{ghazalba@yorku.ca}
\orcid{0002-4715-5371}
\author{Aijun An}
\authornotemark[1]
\email{aan@yorku.ca}
\author{Manos Papagelis}
\authornotemark[1]
\email{papaggel@yorku.ca}
\affiliation{%
  \institution{Department of Electrical Engineering and Computer Science, York University, Toronto, Ontario}
  \streetaddress{}
  \city{Toronto}
  \state{Ontario}
  \country{Canada}
  \postcode{}
}

\renewcommand{\shortauthors}{Babanejad et al.}

\begin{abstract}
Infectious diseases occur when pathogens from other individuals or animals infect a person, resulting in harm to both individuals and society as a whole. The outbreak of such diseases can pose a significant threat to human health. However, early detection and tracking of these outbreaks have the potential to reduce the mortality impact. To address these threats, public health authorities have endeavored to establish comprehensive mechanisms for collecting disease data. Many countries have implemented infectious disease surveillance systems, with the detection of epidemics being a primary objective. The clinical healthcare system, local/state health agencies, federal agencies, academic/professional groups, and collaborating governmental entities all play pivotal roles within this system. Moreover, nowadays, search engines and social media platforms can serve as valuable tools for monitoring disease trends. The Internet and social media have become significant platforms where users share information about their preferences and relationships. This real-time information can be harnessed to gauge the influence of ideas and societal opinions, making it highly useful across various domains and research areas, such as marketing campaigns, financial predictions, and public health, among others. This article provides a review of the existing standard methods developed by researchers for detecting outbreaks using time series data. These methods leverage various data sources, including conventional data sources and social media data or Internet data sources. The review particularly concentrates on works published within the timeframe of 2015 to 2022. 
\end{abstract}




\maketitle

\section{Introduction}

The implementation of automated methods in public health has proven effective in the early detection of naturally occurring outbreaks or those related to bioterrorism. These methods aim to minimize the time between the appearance of strains and the identification of the outbreak. This reduced time gap allows for more efficient investigation and intervention in controlling the disease. Numerous techniques have been developed to detect and forecast outbreaks using routinely collected data.
Surveillance stands as a crucial activity within public health, providing vital information for the protection and promotion of health. It plays a critical role in rapidly identifying disease outbreaks (detection) and in predicting their future development (forecasting), thereby guiding interventions to control epidemics. With concerns surrounding bioterrorist attacks and the emergence of diseases like COVID-19, SARS, and influenza, public health surveillance has become a renewed priority for national security and public well-being.
Many public health agencies now have real-time access to substantial amounts of data from various sources, including clinical settings and telehealth advice centers. While these data hold significant potential for identifying emerging public health threats, their sheer volume and lack of specificity pose new challenges for analysis. To leverage these novel data sources, many public health agencies have implemented automated surveillance systems capable of monitoring data in real-time or near real-time \cite{aakash_forecasting_2021,chan_using_2011,yang_mining_2013, abat2016traditional, haleem2021telemedicine, monaghesh2020role}. These systems have traditionally utilized information sources such as the World Health Organization (WHO), ministries of health, hospital and clinical records, pharmacy records, and laboratory results. In this review article, we refer to these data sources as conventional data sources. However, for early epidemic detection and forecasting, these conventional data sources are less timely and sensitive due to factors such as the long process of data validation, the influence of bureaucracy, politics, higher costs, and resource requirements \cite{chan_using_2011,yang_mining_2013}. The WHO website states that early indicators for more than 60\% of epidemics can be found through informal sources such as social media. Therefore, conventional data can be supplemented with publicly available data from internet-based platforms such as search engines, social media, blogs, or forums \cite{yang_mining_2013,christaki_new_2015,santillana_combining_2015,quincey2009early,al2016potential,xie_detecting_2013,bohlin_tracking_2012,lampos2015advances}. Figure 1 depicts the different types of data sources that have been used for outbreak detection and forecasting in this review article.

\begin{figure}[h]
  \centering
  \includegraphics[width=\linewidth]{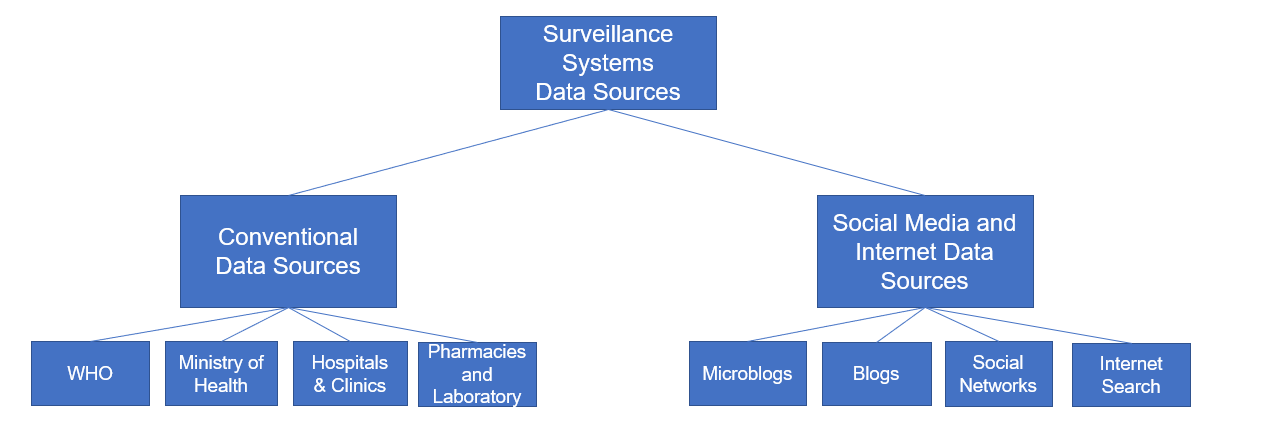}
  \caption{Data sources used for outbreak detection and forecasting}
  \Description{Data sources used for outbreak detection and forecasting}
\end{figure}


Using social media and internet data sources for detecting and forecasting public health events has emerged as a new relevant discipline called Epidemic Intelligence. Epidemic Intelligence Systems (EIS) have been used by public health organizations as monitoring mechanisms for the early detection of disease outbreaks and forecasting their potential spread, which helps reduce the impact of epidemics \cite{linge2009internet, de2010early}. Several notable examples demonstrate the application of Epidemic Intelligence Systems (EIS) for early disease outbreak detection and forecasting. One example is the Google Flu Trends project, developed by Google, which aims to identify flu outbreaks in their early stages by analyzing search queries related to flu symptoms and treatment. By monitoring users' search patterns, the system can provide near real-time estimates of flu activities, enabling prompt responses from public health organizations to potential outbreaks \cite{ginsberg2009detecting}. Another example is the use of Twitter for Disease Surveillance, where researchers utilize Twitter data to monitor and detect disease outbreaks. By analyzing tweets containing keywords related to symptoms or diseases, public health agencies can identify emerging outbreaks and potential hotspots in real-time, allowing for targeted interventions and efficient resource allocation \cite{chew2010pandemics}. Additionally, ProMED-Mail serves as an internet-based reporting system that fosters the sharing and discussion of disease outbreaks and health events among a global network of experts. Acting as an early warning system, ProMED-Mail facilitates the rapid dissemination of information regarding emerging infectious diseases and outbreaks worldwide \cite{madoff2005internet}. Most studies suggest that integrating data from conventional data sources with data from epidemic intelligence systems improves the ability to detect and forecast outbreaks \cite{quincey2009early,al2016potential,xie_detecting_2013,bello-orgaz_survey_2015,al-garadi_using_2016,hornmoen_social_2018}.

Statistical and machine learning techniques have been applied to the prediction, detection, and monitoring of outbreaks using the aforementioned data sources. Since the data from these sources are temporal in nature (that is, having time stamps), the algorithms and methodologies for outbreak and epidemic detection and forecasting are often based on time-series analysis, which can be categorized as shown in Figure 2.

\begin{figure}[h]
  \centering
  \includegraphics[width=\linewidth]{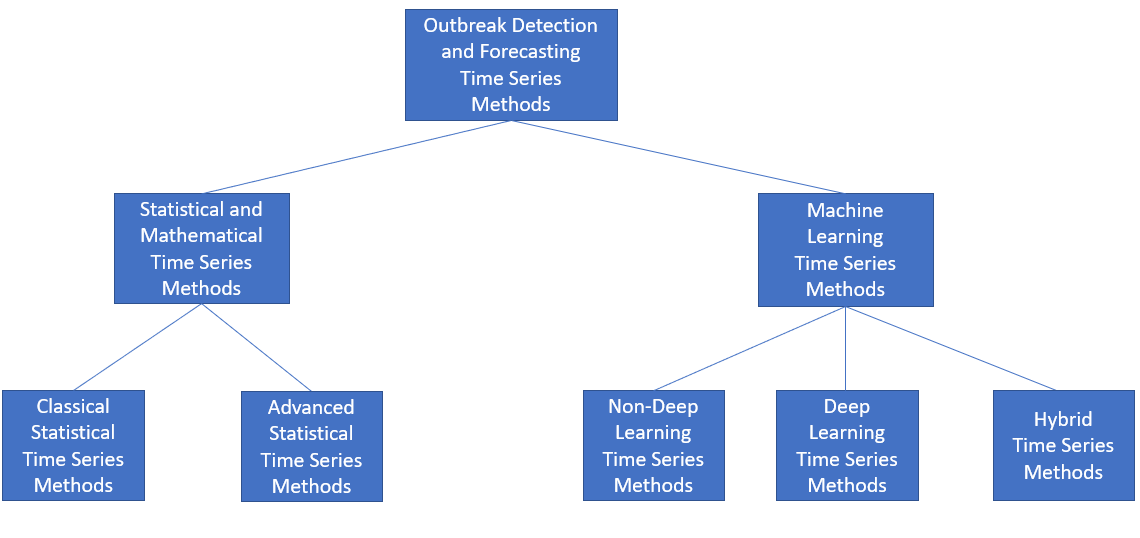}
  \caption{Outbreak detection and forecasting Time Series Methods Categorization}
  \Description{Categorization of Outbreak Detection Time Series Methods}
\end{figure}


The main objectives and contributions of this survey are as follows:

\begin{itemize}

\item To provide a comprehensive overview of the most commonly used outbreak detection and forecasting approaches.

\item To explore the application of both statistical and machine learning methods in disease outbreak detection and forecasting using conventional data sources, as well as social media or internet data.

\item To address the existing literature gap by bringing together the various approaches and methodologies used in outbreak detection and forecasting.

\item To assist academics and researchers interested in this field by providing a summary of recent works including information on methods, limitations, and future works to identify potential future research directions in disease outbreak detection and forecasting.

\item To promote the exchange of ideas and facilitate the development of new approaches and methodologies in disease outbreak detection and forecasting.
\end{itemize}

The focus of this survey is on publications between 2015 and 2022. Similar works will be cited as references to avoid duplication of effort. Below is an overview of how the rest of the article is structured: Section \ref{sec:metods} outlines the procedures followed to complete this article. Section \ref{sec:stats} discusses the statistical and mathematical time series methods employed by health authorities to identify and forecast outbreaks, utilizing data from both conventional data sources and social media or internet sources. Section \ref{sec:ML} explores the role of machine learning time series in outbreak detection and forecasting, specifically focusing on the use of social media and internet data. The challenges encountered during this process, along with potential future research directions, are outlined in Section \ref{sec:challenge}. Finally, Section \ref{sec:conc} provides a summary of all the methods discussed in the article.

\section{Methods}\label{sec:metods}
The purpose of this study is to look into the efficacy and limitations of outbreak detection methods utilizing conventional or social media/Internet data. A systematic literature review was conducted for this research study with the primary goal of summarizing outbreak detection methods and their challenges. The publications were first filtered based on their titles and abstracts. Second, the complete texts were examined. The survey shows the evolution of outbreak detection methods from classical to current methods that use artificial intelligence. 

\subsection{Types of studies}
We study research works that tracked an epidemic using traditional or social media/internet data sources.
Conventional (or non-social media) data sources refer to the data from the World Health Organization (WHO), ministries of health, hospital and clinical records, pharmacy records, and laboratory results, among other sources. Social media/internet data refer to those from a system that allows for the interchange and distribution of information as well as social interaction among individuals and search queries. An outbreak refers to an epidemic that has spread across a geographical area, affecting a significant number of people. 

\subsection{Search strategy to find relevant studies}
We used PUBMED, IEEExplore, ACM Digital Library, Google Scholar, and Web of Science to search the electronic literature for relevant papers with search keywords/phrases of "online social networks," "Healthcare", "Surveillance systems", "Traditional Surveillance systems", "microblogs", "Facebook", "Twitter", "Myspace", "outbreak detection method", "YouTube", "LinkedIn", "Google+", "Friendster", "social media", "social website", "flu", "pandemic", "epidemic", "infectious disease", "Covid-19", "seasonal flu", "H1N1", "HIV", "influenza", "Influenza-like Illnesses", "Ebola", and "Zika." Synonyms and related terms, including case-sensitive variations, were used to generate additional search keywords for each disease. Regardless of the language of the studied data, all English publications were retrieved. No country was barred.

\subsection{Screening period and selection of criteria}
In the initial phase of screening, a total of 3415 articles were collected. Subsequently, the removal of duplicate articles reduced the count to 2274. Upon close examination, only studies aligned with the objectives of our systematic review, as indicated by their titles, were included. These qualifying papers, totaling 384, underwent further scrutiny with specific criteria:

\begin{itemize}
\item The paper must be published in English between 2015 and 2022. 

\item The paper should investigate an outbreak or epidemic and describe the techniques that have been used for detection or monitoring them that extended across a significant geographical region and affected a huge number of people.

\item The study's data source should be derived from any social media, internet, or conventional data sources as explained in the types of studies section.

\end{itemize}

Following this rigorous screening process, 165 articles remained for more comprehensive evaluation. Aalthough the initial search included platforms like YouTube, LinkedIn, and Friendster, no relevant articles using these data sources were found during the final selection.
The next step involved a collaborative effort by the authors to compare and discuss their findings, ultimately leading to a consensus. From this meticulous selection process, a final set of 50 papers emerged for detailed examination. These chosen articles were identified for their exceptional novelty, offering in-depth explanations of innovative methodologies and algorithms. Additionally, some related works were referenced rather than elaborated upon, serving as supporting evidence. Similarly, references were made to certain initial methods and algorithms for context. 

\begin{table*}
\small
  \caption{Classical Statistical Time Series Methods Summary}
  \label{tab:commands}
  \begin{tabular}{p{0.25\linewidth}p{0.7\linewidth}}
    \toprule
    Model & Interpretation\\
    \midrule
    \texttt{Early Aberration Reporting System (EARS) \cite{yang_simulation-based_2018}} & EARS is a statistical surveillance system designed to detect and monitor unusual patterns or aberrations in public health data, such as disease counts or other health-related events. EARS algorithms analyze data over time to identify deviations from expected values, helping to detect potential disease outbreaks or unusual events early and facilitating timely public health responses. \\
    \midrule
    \texttt{Holt-Winters \cite{chatfield_holt-winters_1988,zhu_forecasting_2021,panda_application_2020,yang_simulation-based_2018,hansun_tuned_2021,mahmud_prediction_2021,zhang2017monitoring,djakaria2021covid}}& Holt-Winters is a time series model encompassing three key elements: average value, trend, and seasonality. It combines three simpler smoothing methods—Simple Exponential Smoothing (SES), Holt’s Exponential Smoothing (HES), and a cyclical pattern—to enable forecasting.\\
    \midrule
    \texttt{Holt-Winters Additive Method (HWAAS) \cite{panda_application_2020,hansun_tuned_2021,mahmud_prediction_2021,shukur_time_2021,wickramasinghe_forecasting_2022,fong_finding_2020}}& HWAAS extends Holt's exponential smoothing to include seasonality. It employs exponential smoothing for forecasting level, trend, and seasonal adjustments. Using an additive approach, it combines seasonality with trended forecast, creating the curved Holt-Winters additive forecast. This method is suitable for data with stable trend and seasonality that doesn't grow over time, effectively depicting seasonal fluctuations in the forecast.\\
    \midrule
    \texttt{Moving Average (MA) \cite{earnest_using_2005,dansana_global_2020,pourghasemi_assessment_2020,steiner_detecting_2010}}& MA is defined as an average of a fixed number of items in the time series which move through the series by dropping the top items of the previous averaged group and adding the next in each successive average.\\
    \midrule
    \texttt{Auto-Regressive (AR) \cite{nassar_modeling_2004,khakharia_outbreak_2021,zhang_using_2018,chintalapudi_covid-19_2020}} & AR is a time series model that uses observations from previous time steps as input to a regression equation to predict the value at the next time step.\\
    \midrule
    \texttt{Vector Auto-Regressive (VAR) \cite{monllor_covid-19_2020,rajab_forecasting_2022,oliva_projection_2021,wang_vector_2021,lutkepohl_vector_2013}}& VAR is a multivariate forecasting algorithm that can be used when two or more time series influence each other. It relates current observations of a variable with past observations of itself and past observations of other variables in the system. Model is useful when one is interested in predicting multiple time series variables using a single model. VAR models differ from univariate autoregressive models because they allow feedback to occur between the variables in the model.\\
    \midrule
     \texttt{Cumulated SUM (CUSUM) \cite{yang_simulation-based_2018,watkins_applying_2008,sharifolkashani_early_2021,karami_early_2017}}& The CUSUM control chart is a method for detecting whether the mean of a time series process has shifted beyond some tolerance (i.e., is out-of-control). Originally developed in an industrial process control setting, the CUSUM statistic is typically reset to zero once a process is discovered to be out of control since the industrial process is then recalibrated to be in control. The CUSUM method is also used to detect disease outbreaks in prospective disease surveillance, with a disease outbreak coinciding with an out-of-control process.\\
     \midrule
     \texttt{Auto-Regressive Moving Average (ARMA) \cite{zhu_forecasting_2021,pourghasemi_assessment_2020,buendia_disease_2015,lynch_application_2021}}& ARMA is a model of forecasting in which the methods of autoregression (AR) analysis and moving average (MA) are both applied to time-series data that is well behaved. The AR parameters are first estimated, and then the MA parameters are estimated based on these AR parameters. In ARMA it is assumed that the time series is stationary and when it fluctuates, it does so uniformly around a particular time.\\
      \midrule
     \texttt{Auto-Regressive Integrated Moving Average (ARIMA) \cite{earnest_using_2005,dansana_global_2020,pourghasemi_assessment_2020,singh_prediction_2020,toga_covid-19_2021,arunkumar_forecasting_2021}}& ARIMA model is a combination of the differenced autoregressive model with the moving average model. The AR part of ARIMA shows that the time series is regressed on its own past data. The MA part of ARIMA indicates that the forecast error is a linear combination of past respective errors. The “I” in the ARIMA model stands for integrated; It is a measure of how many non-seasonal differences are needed to achieve stationarity. If no differencing is involved in the model, then it becomes simply an ARMA.\\
    \midrule
     \texttt{Seasonal Autoregressive Integrated Moving Average (SARIMA) \cite{rahmanian2021predicting}}
    &  SARIMA is a mathematical framework used to predict future values of a time series by considering its past values, incorporating differencing to achieve stationarity, and accounting for both non-seasonal and seasonal patterns in the data. It combines autoregressive, integrated, and moving average components with additional seasonal terms.\\
     
    \bottomrule
 \end{tabular}
\end{table*}

\begin{table*}
\small
  \caption{Advanced Statistical Time Series Methods Summary}
  \label{tab:commands}
  \begin{tabular}{p{0.25\linewidth}p{0.7\linewidth}}
    \toprule
    Model & Interpretation\\
    \midrule
 \texttt{Markov switching model (MSM) \cite{amoros_spatio-temporal_2020,salvador_bayesian_2017, lu_bioterrorism_2008}}& A Markov process is one where the probability of being in a particular state is only dependent upon what the state was in the previous period. Transitions between different regimes are governed by fixed probabilities. This model involves multiple structures that can characterize the time series behaviors in different regimes, states or episodes. it is used to describe how data falls into unobserved regimes. Markov models can be an effective way of predicting in time series. The discretization of the state space is of importance for the quality of prediction. Time windows grouped in sequences were used to obtain good transition matrices.\\

 \midrule
 \texttt{Spatio-temporal Bayesian Markov switching model \cite{amoros_spatio-temporal_2020,salvador_bayesian_2017, lu_bioterrorism_2008}}& Spatio-temporal Bayesian Markov switching model is a statistical technique that models data with both spatial and temporal dimensions, allowing for shifts between different states over time and space. It employs Bayesian methods to estimate hidden states and capture transitions between them, making it effective in understanding complex spatio-temporal patterns in various fields such as ecology, epidemiology, and economics.\\

 \midrule
 \texttt{Markov Chain Monte Carlo (MCMC) \cite{bartolucci2022spatio}}& MCMC is a computational technique used in statistics to approximate complex probability distributions. It involves constructing a sequence of correlated samples that converge to the desired distribution, enabling efficient estimation of quantities of interest and uncertainty assessment.\\

 \midrule
 \texttt{Bayesian Structural Time Series (BSTS) \cite{feroze2021analysis}}& BSTS is a statistical framework used for time series modeling and forecasting. It combines Bayesian principles with structural components to capture various patterns, trends, and seasonality in data. This method provides flexible modeling, uncertainty estimation, and is valuable for analyzing complex time series with interpretable results.\\
 
 \bottomrule
 \end{tabular}
\end{table*}

\section{Statistical and Mathematical Time Series Methods}\label{sec:stats}

One popular method for predicting and identifying epidemics is time series analysis. A time series is a series of numerical data in successive order. The time series have been used to track the movement of data, such as stock price, and the number of infected people over a specific period. Many countries conducted infectious disease surveillance in order to detect epidemics at an early stage. This section is organized into two subsections: section 3.1 focuses on the classical statistical time series methods while section 3.2 explains the advanced statistical methods. Table 1 and 2 summarizes some of the statistical and mathematical methods used by authors for outbreak detection using time series data.

\subsection{Classical Statistical Time Series Methods}
Classical statistical methods, like AR, ARMA, ARIMA, VAR, Holt-Winters, and SARIMA, are linear techniques for time series analysis. These methods capture straightforward trends in data and are used with both conventional and social media sources, as detailed in Table 1. They remain valuable for understanding temporal patterns in outbreak detection research.

\subsubsection{Conventional Data Sources}

Ensuring high specificity in epidemic alerts is crucial for infectious disease surveillance. Numerous studies have focused on detecting epidemics \cite{costagliola_routine_1991,snacken_five_1992,stroup_bayesian_1993,choi_evaluation_1981,stroup_evaluation_1993,cliff_changing_1992,nobre_monitoring_1994}, particularly influenza-like illnesses on a national scale \cite{stroup_evaluation_1993,cliff_changing_1992,nobre_monitoring_1994}, while smaller areas, like cities, often receive less attention. Various approaches, including time series analysis, have been used for outbreak detection. This section reviews statistical techniques, with Table 3 summarizing classical methods, their performance compared to health authority data, and future research directions.

\begin{table*}
\small
  \caption{Outbreak Detection Using Classical Statistical Methods and Conventional Data Sources}
  \label{tab:commands}
   \begin{tabular} {p{0.1cm}| p{4cm} p{3cm} p{3cm} p{4cm}} 
    \toprule 
    {\bfseries No} & {\bfseries Models}& {\bfseries Data Source} & {\bfseries Result} & {\bfseries Limitations/Future work} \\
    \midrule

 \multirow{ 8}{*}{ 1}    
    &
    \begin{itemize}
        \item Auto-Regressive Moving Average(ARMA) \cite{buendia_disease_2015} (2015)
    \end{itemize}
   & Philippines health and health-related agencies
    & ARMA helped to know how the 
    outbreak will turn out.&
   Future work could involve adapting the system for compatibility with various epidemiological models and integrating Geographic Information System (GIS) data to offer location-specific insights into factors influencing disease spread.\\
    \midrule
    \multirow{ 8}{*}{ 2}  
    & 
\begin{itemize} 
    \item CUSUM(glm with trend) 
    \item CUSUM(standard), CUSUM(rossi), CUSUM(rossi and glm with trend), EARS C1,C2,C3
, ARIMA, Holt-Winters
    \cite{yang_simulation-based_2018} (2018)
 \end{itemize}

    &
    
    Time series generated including trends, seasonality, and randomly occurring Outbreaks, and real-world daily and weekly data related to diarrhea infection. &
    Based on the findings, the authors proposed sMAPE evaluation metrics for assessing the performance of syndromic surveillance analysis when the data lacks the outbreak state variable, and they demonstrated that the "glm with trend variable" CUSUM algorithm outperforms other default CUSUM algorithms. &  
    Not Available \\
    \bottomrule

    \multirow{ 8}{*}{ 3}  
    & 
\begin{itemize}
    \item ARIMA \cite{roy2021spatial} (2021)
     \end{itemize}
    &
    Official Indian State Health Offices, 
Geospatial data included district boundaries
    &
    Results highlighted the effectiveness of the ARIMA model in accurately predicting the incidence of COVID-19 cases, aiding in epidemiological surveillance and providing insights into high-risk regions.
    &
   The limitations are:

1. It relies on reported case data, which might underestimate actual cases; 2. It only considered linear relationships, limiting capturing complex dynamics.
3. It assumes stationarity in data, ignoring potential seasonality.
4. Generalization might be limited due to the specific disease context. \\
\midrule
\multirow{ 7}{*}{ 4}  
    & \begin{itemize}
 \item VAR \cite{wang_vector_2021} (2021)
 \end{itemize}
 & 
 COVID-19 disease from CDC (Centers for Disease Control and Prevention) in
fifty states of the United States by the Center for Systems Science and Engineering (CSSE) at Johns Hopkins University
 & 
 The VAR model for this dataset can predict the number of daily positive cases with high accuracy for the next 30 days based on the past 8 days.
 &
 Future work involves incorporating relevant variables such as social distancing measures and vaccination rates to enhance the model's performance, especially in the presence of external factors affecting viral transmission. This would extend the model's utility beyond the vaccine rollout, facilitating predictions of infection reduction during the initial vaccination phase.\\
    \bottomrule
 \end{tabular}
\end{table*}

Referring to Table 3, the initial approach, pioneered by Buendia and Solano in 2015 \cite{buendia_disease_2015}, introduces a Disease Outbreak Detection System. This online system serves as a tool for aiding public health professionals in identifying and monitoring disease outbreaks. Employing the Autoregressive Moving Averages (ARMA) model, it gathers health-related data from surveys, censuses, and administrative records of health agencies in the Philippines. By analyzing this dataset, the system generates predictive values for specific time intervals. Consequently, epidemiologists gain the ability to foresee the progression of outbreaks and implement necessary measures for containment and resolution.

The subsequent method, detailed by Yang et al. in 2018 \cite{yang_simulation-based_2018}, focuses on a comprehensive evaluation of diverse techniques to enhance the current disease outbreak detection system at the Korea Centers for Disease Control and Prevention (KCDC). Through a meticulous comparative study, the CUmulative SUM (CUSUM), Early Aberration Reporting System (EARS), autoregressive integrated moving average (ARIMA), and Holt-Winters algorithm are assessed for temporal outbreak detection. This scrutiny involves a wide range of time series, encompassing trends, seasonality, and sporadic outbreaks, coupled with real-world daily and weekly data pertaining to cases of diarrhea infection. The evaluation employs a multitude of metrics, including sensitivity, specificity, positive predictive value, negative predictive value, F1 score, symmetric mean absolute percent error, root-mean-square error, and mean absolute deviation. These metrics collectively offer insights into the algorithms' performance.

In the context of this comparison, the EARS C3 method emerges as superior to the other algorithms scrutinized in the study. However, the Holt-Winters algorithm excels when both baseline frequency and dispersion parameter values are below 1.5 and 2, respectively. This research underscores the significance of algorithmic performance and judicious metric selection, as these elements intricately correspond to the data's characteristics concerning trends, seasonality, and baseline infections.

Both of these research endeavors contribute invaluable insights to the domain of disease outbreak detection. Buendia and Solano's work furnishes a pragmatic system for tracking outbreaks, while Yang et al.'s study presents a thorough analysis of distinct algorithms, striving to refine outbreak detection precision. These investigations address distinct facets of outbreak detection systems, collectively emphasizing the pivotal roles of appropriate models and metrics in the realm of public health decision-making.

The third scholarly endeavor, conducted by Roy et al. (2021) \cite{roy2021spatial}, is predominantly centered around surmounting the challenges posed by the COVID-19 pandemic. The study's core objectives encompass the development of effective short-term prediction models and the execution of spatial analyses to unravel disease distribution patterns. Leveraging data amassed from January to May 2020, the researchers harness Geographic Information System (GIS) techniques to gauge disease risks across Indian districts. Significantly, they employ the Autoregressive Integrated Moving Average (ARIMA) model for time-series forecasting, with a particular focus on cumulative confirmed COVID-19 cases in states witnessing high daily incidence rates. This endeavor contributes substantially to pandemic preparedness by harnessing statistical models to enhance predictive precision and spatial comprehension.

The concluding review examines the work of Wang et al. (2021) \cite{wang_vector_2021}, where a generalized VAR model is harnessed to prognosticate the dynamics of COVID-19 cases. Utilizing a VAR model enables the capture of dynamic linear correlations between variables that mutually influence one another. The model incorporates a range of correlated factors, encompassing undetected infections, reported deaths, and environmental variables. The authors propose that this modeling approach can be extended to forecast other epidemics characterized by COVID-19-like attributes.

\subsubsection{Social Media and Internet Based  Data Sources}

The internet's rise offers new opportunities for delivering critical health information swiftly, as highlighted by Al-Shorbaji \cite{al2016potential}. Unlike traditional methods, which were often slow, web-based platforms enable rapid dissemination and analysis. Social media and online searches allow hospital networks to engage patient advocacy groups in real time, as noted by Thaker et al. \cite{thaker_how_2011}, while also facilitating information exchange at health conferences. Additionally, these platforms provide channels for public engagement and enhance patient-provider communication.

\begin{table*}
\small
  \caption{Outbreak Detection Using Classical Statistical Methods and Social Media or Internet Data Sources}
  \label{tab:commands}
   \begin{tabular} {p{0.2cm}| p{3cm} p{3cm} p{3cm} p{4cm}} 
    \toprule 
    {\bfseries No} & {\bfseries Models}& {\bfseries Data Source} & {\bfseries Result} & {\bfseries Limitations/Future work} \\
    \midrule
 \multirow{ 5}{*}{ 1}  
& 
\begin{itemize}
 \item SARIMA
 \item Regression Tree Model
 \cite{chen2019avian} (2019)
\end{itemize}
&
   Laboratory-confirmed H7N9 cases reported in China (2013-2017),
Baidu Search Index (BSI) for keywords "H7N9," "Avian influenza," and "Live poultry",
Weibo Posting Index (WPI) for the same keywords  &
The Seasonal Autoregressive Integrated Moving Average (SARIMA) model, the Cross-Correlation Analysis, and the Regression Tree Model, all of which used BSI and WPI data to predict H7N9 cases and suggest their potential as early warning indicators for disease outbreaks.
     & 
    Limitations include using provincial capital coordinates for spatial analysis and lacking access to detailed geographic distribution of Weibo Post Index (WPI) data.
    
    Future research directions in tracking disease patterns on various social media platforms and analyzing human attitudes or reactions toward health hazards using content analysis of social network posts.\\

\midrule
 \multirow{ 8}{*}{ 2}    
    &
    \begin{itemize}
        \item ARIMA(X)
        \item Custom Approximate Model
        \cite{samaras_comparing_2020} (2020)
    \end{itemize}
   & Data on influenza in Greece have been collected from Google and Twitter
    & Although the alternative model's results are also reliable for outbreak detection, the ARIMA(X) model outperforms it. Twitter data is also slightly better than Google data for ARIMA (X) model results.&
    Limitations include the data size and electronic files were small, and Greek language peculiarities made tweet localization unnecessary. However, tracking tweets based solely on language, such as "influenza" in English, may require locating tweets to study specific areas. Another limitation is the use of REST APIs, which have usage restrictions known as rate limits, impacting each user or application within 15-minute windows.
\\
    \midrule
    \multirow{ 5}{*}{ 3}  
& 
\begin{itemize}
 \item SMSI
 \cite{qin_prediction_2020} (2020)
\end{itemize}
&
    Baidu Search Index in Social Media  &
    SMSI could be used to predict COVID-19 outbreaks in affected populations, and it has a high correlation with new suspected and confirmed COVID-19 infection cases. & 
    Not Available \\

    \bottomrule
 \end{tabular}
\end{table*}

Table 4 provides insight into two recent studies that harnessed internet data for outbreak detection, juxtaposing these approaches with real-world data while also presenting their limitations and avenues for future exploration. Chen et al.'s 2019 study \cite{chen2019avian} delves into the potential of leveraging internet search queries and social media data to detect and monitor avian influenza A (H7N9) cases in China. This investigation delves into the spatial and temporal trends of H7N9 cases alongside related internet search queries. The analysis reveals positive correlations between H7N9 cases and Baidu Search Index (BSI) and Weibo Posting Index (WPI) data, indicating early warning potential. Utilizing models like SARIMA and regression trees, the study predicts H7N9 cases based on search engine and social media data, demonstrating predictive prowess and highlighting the role of mobile access to health information.

Interestingly, both BSI and WPI exhibit temporal and spatial consistency with H7N9 cases, offering potential precursors to outbreak trends. SARIMA models underscore BSI's superiority over WPI in terms of sensitivity and specificity. Regression tree analysis identifies key predictors of H7N9 occurrence: BSI with a lag of 0 weeks and WPI with a lag of -1 week.

In 2020, Samaras et al. \cite{samaras_comparing_2020} developed a system for detecting and predicting severe epidemics using data from search engines and social networks. The study compared Twitter and Google for tracking influenza in Greece, using an ARIMA(X) model on weekly data and a custom model on daily data, against official EU statistics. The research aimed to evaluate the suitability of these platforms for epidemic tracking and their predictive capabilities. It found that Twitter outperformed Google in tracking influenza, with the ARIMA(X) model proving superior, though both models were reliable.

Lastly, Qin et al.'s research in 2020 \cite{qin_prediction_2020} delved into the connection between new COVID-19 cases and Baidu search index (BSI) data from a prominent Chinese social network. Their aim was to develop an affordable and effective model for predicting new COVID-19 cases, aiding in timely policy decisions. Employing various methods for coefficient estimation, the study identified a strong correlation between new suspected COVID-19 case numbers and the lagged series of the social media search index. Remarkably, the social media search indexes (SMSI) findings anticipated new COVID-19 cases 6-9 days in advance. The subset selection method was optimal, providing low estimation error and a moderate number of predictors, and the SMSI findings closely correlated with newly confirmed COVID-19 cases.

Both research endeavors harnessed internet data, particularly social media and search engine information, to predict and monitor epidemic outbreaks. The first study emphasized Twitter's superiority over Google for influenza tracking, while the second study focused on the Baidu search index's correlation with new COVID-19 cases. Methodologies varied across the studies, with the first employing ARIMA(X) and a custom model, and the second exploring five different coefficient estimation methods. Collectively, these investigations underscore the potential of internet data in forecasting and monitoring epidemic outbreaks, thereby enabling timely public health interventions.

\subsection{Advanced Statistical Time Series Methods}
Advanced Statistical Methods extend beyond linear models to capture complex dynamics and uncertainties. These include the Spatio-temporal Bayesian Markov Switching Model, Markov Switching Model (MSM), Markov Chain Monte Carlo (MCMC), and Bayesian Structural Time Series (BSTS), which model non-linear relationships and quantify uncertainty using Bayesian principles. These methods are ideal for capturing nuanced temporal behaviors and making predictions in intricate time series. Table 2 summarizes prominent advanced statistical methods used for outbreak detection with conventional and social media data sources.

\subsubsection{Conventional Data Sources}

In this section, we delve into studies that have harnessed advanced statistical outbreak detection methods while relying on conventional data sources rather than social media inputs. As detailed in Table 5, the first review centers on a study by Rahmanian et al. (2021) \cite{rahmanian2021predicting}. This research aimed to investigate the potential influence of environmental variables on cutaneous leishmaniasis occurrences, employing time-series models. A key objective was comparing the predictive prowess of seasonal autoregressive integrated moving average (SARIMA) models with the Markov switching model (MSM). Leveraging yearly and monthly data spanning from January 2000 to December 2019, the study encompassed 49,364 confirmed cases of cutaneous leishmaniasis in Isfahan province, Iran. Data on humidity, wind speed, and vegetation were sourced from the Leishmaniasis National Surveillance System, the Isfahan Province Meteorological Organization, and the Iranian Space Agency.

Findings unveiled significant associations between cutaneous leishmaniasis outbreaks and various environmental factors at varying time lags, particularly minimum and maximum relative humidity alongside wind speed. The comparative analysis of SARIMA and MSM models highlighted the latter's superiority in metrics like Akaikes information criterion (AIC) and mean absolute percentage error (MAPE). This study underscores the efficacy of both SARIMA and MSM models for cutaneous leishmaniasis prediction, with the MSM approach emerging as a recommendation due to its dynamic nature and insightful potential in comparison to single-distribution models.

In a distinct endeavor, Feroze et al. (2021) \cite{feroze2021analysis} navigated the exigent landscape of Pakistan's COVID-19 pandemic. This research aimed to analyze and predict disease trends, essential given the strain on the country's healthcare infrastructure. Bayesian structural time series (BSTS) models were employed to gain a nuanced grasp of the pandemic's trajectory over the ensuing 30 days. Of particular note, the study introduced a unique dimension by probing the causal impacts of lockdown lifting through intervention analysis within the BSTS framework.

BSTS models emerged as pivotal tools, enabling a comprehensive assessment of pandemic trends. The authors thoughtfully contrasted these models with the commonly used autoregressive integrated moving average (ARIMA) models, underscoring BSTS models' advantages in assimilating prior information, accommodating covariates, and evolving over time. By applying BSTS models, the study illuminated Pakistan's pandemic trajectory, predicting exponential case growth paired with an optimistic trend of swifter recovery than new case emergence. This granular understanding informed effective interventions, hotspot identification, and adherence to Standard Operating Procedures (SOPs), tailored to Pakistan's distinct economic and healthcare landscape.

Importantly, this research's impact extended beyond Pakistan's borders, providing insights into neighboring countries such as Iran and India. Through this comparative lens, decision-makers and healthcare professionals gleaned valuable guidance. While acknowledging data-related limitations and sustained trend assumptions, the study furnishes a robust framework for pandemic dynamics analysis and projection.

Transitioning to another study, Bartolucci and Farcomeni's work \cite{bartolucci2022spatio} introduces the Discrete Latent Variable Model for COVID-19, a spatio-temporal model tailored for analyzing SARS-CoV-2 incident cases. This model integrates discrete latent variables evolving over time in a Markov chain, capturing spatial dependencies among neighboring regions. Employing Poisson regression, the model considers a common trend modulated by the latent state, influenced by environmental variables. The analysis, conducted using Italian regional COVID-19 data, unveiled distinct risk profiles across time and space, effectively categorizing areas based on infection severity levels.

The model effectively accounts for spatial relationships, incorporates the number of swabs as an offset to mitigate bias, and offers insights into regional risk patterns. This study harnesses the model's flexibility to characterize different pandemic phases and intervention impacts, providing a deeper understanding of regional trends.

Shifting focus to Thorakkattle et al.'s research \cite{navas2022forecasting}, Bayesian structural time series (BSTS) models were employed to forecast COVID-19 trends and assess vaccination's causal effects in multiple countries. The study aimed to furnish more adaptable and accurate predictions than traditional ARIMA models. Notably, the BSTS models demonstrated superior accuracy in predicting future COVID-19 cases and deaths, showcasing their efficacy. The research underscores that effective vaccination efforts in the United States, the United Kingdom, and the United Arab Emirates led to reduced mortality, while the impact on case and death rates in India remained limited. The study's insights informed policymakers on the need for prompt vaccination and provided invaluable insights for guiding public health strategies.

Utilizing BSTS models, the research delved into COVID-19 temporal patterns and vaccination's intervention effects, revealing disparities across countries. The study's contributions extend beyond individual nations, with cross-country comparisons informing effective response strategies. Acknowledging the research's limitations related to data reporting and trend assumptions, it offers a robust foundation for analyzing and forecasting pandemic dynamics.

Lastly, Yen et al.'s study \cite{yen2022new} presents a pioneering surveillance approach aimed at predicting community-acquired outbreaks originating from imported cases of new SARS-CoV-2 variants. The study introduces metrics to estimate domestic cluster infection risk and establishes alert thresholds for targeted containment. Employing Bayesian Monte Carlo Markov Chain techniques and extra-Poisson regression models, the research underscores the metrics' value in preventing outbreaks during certain periods and emphasizing their significance for adapting to emerging variants and mitigating large-scale outbreaks.

\begin{table*}
\small
  \caption{Outbreak Detection Using Advanced Statistical Methods and Conventional Data Sources}
  \label{tab:commands}
   \begin{tabular} {p{0.2cm}| p{3cm} p{3cm} p{4cm} p{4cm}} 
    \toprule 
    {\bfseries No} & {\bfseries Models}& {\bfseries Data Source} & {\bfseries Result} & {\bfseries Limitations/Future work} \\
    \midrule
\multirow{ 8}{*}{ 1}    
    &
    \begin{itemize}
        \item Markov Switching Model (MSM)
        \item Seasonal Autoregressive Integrated Moving Average (SARIMA)
        \cite{rahmanian2021predicting} (2021)
    \end{itemize}
   & Yearly and monthly data of 49 364 parasitologically-confirmed cases of CL in Isfahan province 
   &
   The study found significant associations between cutaneous leishmaniasis outbreaks and various environmental factors, such as humidity and wind speed, using both seasonal autoregressive integrated moving average (SARIMA) and Markov switching model (MSM), with the MSM showing improved predictive performance.
    & Limitations include the omission of important parameters like host-related factors, vectors, parasite type, and health interventions, which are crucial for understanding the epidemiology of cutaneous leishmaniasis in the region and should be considered in future research.\\
\midrule
\multirow{ 8}{*}{ 2}    
    &
    \begin{itemize}
        \item Bayesian structural time series (BSTS)
        \item Autoregressive Integrated Moving Average (ARIMA)
        \cite{feroze2021analysis} (2021)
        \end{itemize}
        & 
        NIH Islamabad, Pakistan, and Our World in Data: For Iran and India
        &
        The BSTS models demonstrated higher forecast accuracy than ARIMA models, contributing to robust predictions.  Iran's situation was controlled, while India faced a significant surge.
        &
        Limitations: Underreported data due to limited testing, assumptions of current trends continuing, and lack of investigation into risk factors.
        
        Future Work: Further study on risk factors, incorporating demographic and social network data, refining forecasting models with more comprehensive testing data.\\
    \midrule    
\multirow{ 8}{*}{ 3}    
    &
    \begin{itemize}
        \item Latent Markov model
        \cite{bartolucci2022spatio} (2022)
        \end{itemize}
        & 
        COVID-19 data from the Italian Civil Protection Department.
        &
        The study's analysis of Italian COVID-19 data revealed distinct risk profiles and varying trajectories across regions, capturing the pandemic's dynamics and severity using a Discrete Latent Variable Model.
        &
        Limitation: Assumes spatial dependence based on sharing land borders, and potential for biased spatial structure.

        Future Work: Extend to time-varying latent states, explore more flexible count assumptions, include covariates for interventions, and apply to other disease mapping scenarios.\\
    \midrule   

    \multirow{6}{*}{ 4}    
    &
    \begin{itemize}
        \item BSTS
        \item ARIMA
        \cite{navas2022forecasting} (2022)
        \end{itemize}
        & 
        "Our World in Data" and Humanitarian Data Exchange.
        &
        Bayesian structural time series (BSTS) models revealed more accurate COVID-19 predictions and assessed the causal impact of vaccinations, indicating successful mortality reduction in the United States and the United Kingdom, while India faces challenges due to slower immunization.
        &
        Limitations: Data underreporting, lack of risk factor evaluation, and inherent uncertainty in forecasts due to data quality.
        
        Future Work: Further research on vaccine distribution strategies, incorporating risk factors for improved predictions, and addressing uncertainties in data reporting for enhanced forecasting accuracy.
    \\
    \midrule   

    \multirow{ 5}{*}{ 5}    
    &
    \begin{itemize}
        \item Bayesian MCMC
        \item extra-Poisson regression        \cite{yen2022new} (2022)
    \end{itemize}
   & 
  Taiwanese COVID-19 weekly data
   &
  New surveillance metrics were developed to predict domestic cluster infections and guide containment measures against emerging SARS-CoV-2 variants.
    & 
    Not Available
    \\
\midrule
 \end{tabular}
\end{table*}

\begin{table*}
\small
  \caption{Outbreak Detection Using Advanced Statistical Methods and Social Media or Internet Data Sources}
  \label{tab:commands}
   \begin{tabular} {p{0.2cm}| p{3cm} p{3cm} p{3cm} p{4cm}} 
    \toprule 
    {\bfseries No} & {\bfseries Models}& {\bfseries Data Source} & {\bfseries Result} & {\bfseries Limitations/Future work} \\
    \midrule

 \multirow{ 5}{*}{ 1}    
    &
    \begin{itemize}
        \item SVM
        \item ETAS
        \cite{hassan2019social} (2019)
    \end{itemize}
   & Flu-Related Tweets, 
Cerner Health Fact Clinical Encounter Records
    & 
    The potential of location-based social media data, integrated with clinical records and advanced modeling techniques, to enhance early detection and response to flu outbreaks.
    &
    Limitations of the work include: Incomplete geolocation data from social media users.
    Potential noise and biases in social media data.
    Aggregation issues due to varying spatial units.
    Legal, political, and economic obstacles in utilizing big data.
    Dependence on the quality of data analyzed for accurate insights.
    \\
    \midrule

\multirow{ 8}{*}{ 2}    
    &
    \begin{itemize}
    \item  Spatio-temporal Bayesian Markov switching model
    \cite{amoros2020spatio} (2020)
    \end{itemize}
&
USA Google Flu Trends database
&  
The proposed spatio-temporal Bayesian Markov switching model demonstrated superior performance in early detection of influenza outbreaks compared to variations of the same model without structured spatial components and a purely temporal model for outbreaks detection.
&
Not Available
\\
\midrule
    
    \multirow{ 8}{*}{ 3}  
    & 
\begin{itemize}
    \item Bayesian hierarchical models
    \item MCMC
    \item BSTM
    \cite{wang2021bayesian} (2021)
 \end{itemize}

    &
    Area-level mobile phone data.
Traffic Analysis Zones (TAZs) demographics and mobile phone user counts.
Cell tower locations and user counts.
Spatial and temporal population distribution data for Shenzhen.
    &
    The study effectively utilized a Bayesian spatio-temporal model with area-level mobile phone data to enhance understanding of intra-urban population distribution dynamics.
    &
    Limitations:
Data Privacy Concerns,
Spatial and Temporal Resolution Constraints

Future Work:
Incorporating Explanatory Variables,
Comparing Different Time Periods,
Exploring Varying Effects of Mixed-Use and Urban Dynamics
    \\
    \bottomrule
    
\multirow{ 8}{*}{ 4}    
    &
    \begin{itemize}
    \item Markov Chains
    \cite{suryaningrat2021posted} (2021)
    \end{itemize}
&
Twitter data (crawled with keywords: Coronavirus, covid-19, Covid19)
&  
The study utilized Markov Chains to analyze Twitter data related to COVID-19, revealing a sustained high level of discussion on the topic and suggesting its potential as an information center for the pandemic.
&
Limitations include that it is limited to Twitter data, and may not represent entire public sentiment.
Did not consider external factors influencing discussion patterns.

Future work could involve integrating other data sources and refining the Markov Chain model for improved accuracy.
\\
\midrule

\multirow{ 8}{*}{ 5}    
    &
    \begin{itemize}
        \item HWF
        \item Naïve Bayes classifier
        \item Markov chain
        \item Decision tree
        \cite{pradeepa2021epidemic} (2021)
    \end{itemize}
   & 
   From Twitter with keywords related to COVID-19
   &
  The proposed hypergraph-based technique accurately predicted Twitter users' locations based on COVID-19 content, outperforming other algorithms, and can aid in tracking epidemic zones.
    & 
    Limitations:
Reliance on publicly available data.
Accuracy linked to local words in tweets.

Future Work:
Extend the framework to multiple networks.
Address geolocation issues in different platforms.
    \\
\midrule
 \end{tabular}
\end{table*}

\subsubsection{Social Media and Internet Based Data Sources}

In this section, we shed light on researchers who have harnessed advanced statistical techniques for detecting outbreaks using data derived from social media sources. As outlined in Table 6, the first study, led by Zadeh et al. (2019) \cite{hassan2019social}, adopts a comprehensive approach, melding big data analytics and social media platforms to achieve accurate and timely tracking of flu outbreaks. By amalgamating two primary datasets – flu-related tweets from social media and clinical flu encounter records – this study unfolds the potential of location-based social media platforms for real-time disease surveillance. To ensure data authenticity, a Support Vector Machine (SVM) classifier segregates flu-indicative tweets from non-indicative ones. This classification process enhances dataset accuracy, setting the stage for subsequent analyses.

The integrated dataset substantiates the promise of social media for real-time disease surveillance. By employing a spatio-temporal point processes model rooted in the epidemic-type aftershock Sequence (ETAS) model, the researchers unveil temporal and spatial relationships between online flu-related conversations and actual flu cases. Remarkably, the study unveils that Twitter discussions often anticipate clinical flu encounters by approximately a month, underscoring the predictive potential of social media in tracking disease trends.

This research resonates with the synergy between big data analytics, machine learning techniques like SVM, and intricate spatio-temporal point process models. The fusion of datasets and utilization of sophisticated analytical models not only validates the flu outbreak nowcasting efficacy but also exemplifies potential in enhancing public health interventions and response strategies.

Transitioning to the second review, Amorós et al. (2020) \cite{amoros2020spatio} present a pioneering spatio-temporal Bayesian Markov switching model for swift influenza outbreak detection. This innovative approach strategically incorporates differentiated incidence rates, weaving temporal autoregressive and spatial conditional autoregressive components to capture the intricate spatio-temporal evolution of influenza data. The model's sophistication is evident as it enhances sensitivity, enabling early epidemic peak identification even when initial rates are low. The model's superiority over purely temporal methods materializes through its application to the USA Google Flu Trends database, unveiling improved outbreak detection accuracy and adeptness in handling simultaneous outbreaks originating from distinct time points.

Addressing the pressing need for early influenza outbreak detection, this research elegantly intertwines temporal and spatial methodologies, bolstering surveillance systems. The incorporation of latent variables that classify observations as epidemic or endemic empowers the model to identify outbreaks with varied onsets and durations. By expertly capturing the dynamic nature of influenza spread, particularly through the application of spatial structures, the study delivers a robust framework for prompt and precise influenza outbreak identification. This framework lends substantial potency to public health interventions.

Next, Wang et al.'s study in 2021 \cite{wang2021bayesian} hones in on comprehending population distribution via area-level mobile phone data, employing Bayesian hierarchical models to illuminate space-time patterns. Focused on Shenzhen, China, the research integrates spatial patterns, temporal trends, and deviations through Markov chain Monte Carlo simulation. The model adeptly identifies areas with unstable population trends, thereby informing urban planning and disease response.

By employing Bayesian hierarchical models, the study unravels intricate space-time patterns in intra-urban population distribution. This dual-dimensional approach successfully disentangles predictable trends from unstable fluctuations, enabling the identification of areas with consistent or erratic population changes. The model's incorporation of spatially correlated and uncorrelated random effects augments local pattern comprehension. Future directions encompass exploring alternative spatial priors, integrating explanatory variables' effects, comparing varying temporal periods, and delving into differing coefficients for spatial and temporal influences on population fluctuations. This research contributes substantively to the understanding of urban dynamics and their interplay with space-time patterns, pivotal for efficient urban planning and resource allocation.

The subsequent study, conducted by Suryaningrat et al. (2021) \cite{suryaningrat2021posted}, delves into social media's impact in information dissemination, particularly within data mining research. The study investigates the application of Markov Chains to predict the trajectory of COVID-19 discussions on Twitter. Utilizing tweet data collected over distinct observation periods, the research underscores sustained high levels of COVID-19 conversation on Twitter, suggesting its potential as an information hub. The study encourages further refinement by incorporating more comprehensive data and extended observation periods to enhance the Markov Chain model's precision. This enhancement could empower policymakers in utilizing Twitter as a credible COVID-19 information source.

This study delves into the role of social media, notably Twitter, in disseminating and shaping COVID-19 discussions. Leveraging Markov Chains, the research endeavors to predict the trajectory of Twitter discourse surrounding the pandemic. The study encompasses multiple stages, including data collection, cleaning, processing, and the application of the Markov Chain model. The outcomes underscore the sustained nature of COVID-19 discourse on Twitter, indicating the platform's significance as a pandemic information conduit. The research lays a roadmap for future improvement, advocating the integration of more expansive data and longer observation periods to enhance the Markov Chain model's predictive prowess, offering invaluable insights to policymakers relying on Twitter for COVID-19 updates.

The fifth review by Pradeepa (2021) \cite{pradeepa2021epidemic} delves into detecting COVID-19's geographic spread through the analysis of Twitter user content. Addressing the challenge of undisclosed user locations, the research introduces a hypergraph-based technique, employing weighting factors termed hypergraph with weighting factor (HWF), to predict users' locations. By associating words with locations in a hypergraph, this model boosts location prediction accuracy, thus aiding epidemic zone identification. The technique's superiority over existing methods substantiates its potential for epidemic forecasting and disaster management applications.

This study embarks on analyzing Twitter user content to deduce the geographic spread of COVID-19. The innovative hypergraph-based technique, enhanced by weighting factors, strives to predict users' locations despite the challenge of undisclosed information. Through a multi-step process, the model refines location predictions by mapping words to locations in a hypergraph. The model's efficacy surpasses other methods, spotlighting its potential for forecasting epidemics and informing disaster management strategies. Despite its effectiveness, the technique remains computationally feasible for real-time applications, paving the way for a promising avenue of precise location prediction in an evolving epidemic landscape.

\section{Machine Learning Time Series Methods}
\label{sec:ML}

Machine learning algorithms have the potential to analyze diverse datasets, encompassing information about known viruses, animal populations, human demographics, biology, biodiversity, physical infrastructures, cultural/social practices worldwide, and disease geolocation, to effectively predict disease outbreaks. This section is organized into three subsections: section 4.1 focuses on Non-Deep Learning Time Series approaches, section 4.2 reviews Deep Learning Time Series methods, and section 4.3 covers some hybrid time series models. Each part is further categorized into models using conventional datasets and social media data. Tables 7, 10, and 12 provide concise summaries of common methods and their related definitions.

\subsection{Non-Deep Learning Time Series Methods} 
Table 7 highlights some of the most common non-deep learning algorithms used for outbreak detection, utilizing both conventional and social media time series data sources. The purpose of this table is to offer a concise summary of each algorithm, making it easier for readers to comprehend prior works.

\begin{table*}
\small
  \caption{Non-Deep Learning Time Series Methods Summary}
  \label{tab:commands}
  \begin{tabular}{p{0.25\linewidth}p{0.7\linewidth}}
    \toprule
    Model & Interpretation\\
    \midrule
    \texttt{Least Absolute Shrinkage and Selection Operator (LASSO) \cite{shi_three-month_2016,tang_review_2019,rustam_covid-19_2020,guo_developing_2017,lee_using_2014}}& LASSO is a statistical formula whose main purpose is the feature selection and regularization of data models. It is based on minimizing Mean Squared Error, which is based on balancing the opposing factors of bias and variance to build the most predictive model. It is usually used in machine learning for the selection of a subset of variables. It can find patterns within large datasets while avoiding the problem of over-fitting.\\
    \midrule
    \texttt{Least-Angle Regression (LARS) \cite{efron_least_2004,wang_empirical_2021}}& LARS is a variable selection method with proven performance for cross-sectional data. It is used in regression for high-dimensional data (i.e., data with a large number of attributes). it is extended to time series forecasting with many predictors.\\
    \midrule
     \texttt{XG-Boost \cite{chen2015xgboost,chen_xgboost_2016,jeon_predicting_2020,aakash_forecasting_2021,shashvat_epidemiology_2020,saif_analysis_2022,lalli_optimized_2021,kalipe_predicting_2018,fang_application_2022,badkundri_forecasting_2019}}& XGBoost is an implementation of the gradient boosting ensemble algorithm for classification and regression. XGBoost uses the ensemble of weak prediction models, gradient boosting helps us in making predictions. Examples of weak models can be decision trees. It helps in generalizing the other model by optimizing the arbitrary differential loss function. Time series datasets can be transformed into supervised learning using a sliding-window representation for  XG-Boost.\\
     \midrule
     \texttt{Support Vector Machine (SVM) \cite{singh_prediction_2020-1,gupta_prediction_2021,jakkula2006tutorial}}& SVM, a supervised machine learning algorithm, classifies and analyzes data for classification and regression. It categorizes data, creating wide margins between categories. SVM finds applications in text, images, and time series forecasting. In time series, it maps data to a higher dimension for separation, and for regression, it forecasts nonlinear, non-stationary data with undefined processes.\\
     \midrule
     \texttt{Decision Trees \cite{comert2020malaria,krishnan_predicting_2022,khan_dengue_2022,leopord2016survey,lowie_decision_2021,podgorelec_decision_2002}}& Decision Trees are a type of Supervised Machine Learning (that is you explain what the input is and what the corresponding output is in the training data) where the data is continuously split according to a certain parameter. The tree can be explained by two entities, namely decision nodes, and leaves.\\
     \midrule
     \texttt{Gaussian Naïve Bayes (GNB) \cite{ali_feature-driven_2019,zakiyyah_prediction_2021,zhao_combining_2021,rezaeijo_screening_2021,omadlao_machine_2022}}& Naïve Bayes is a generative model. (Gaussian) Naïve Bayes assumes that each class follows a Gaussian distribution. The difference between QDA and (Gaussian) Naïve Bayes is that Naïve Bayes assumes independence of the features, which means the covariance matrices are diagonal matrices.\\
      \midrule
     \texttt{Random Forest \cite{kane_comparison_2014,liang_prediction_2020,dansana_covid-19_2022,yesilkanat_spatio-temporal_2020,ong_mapping_2018}}& Random Forest is a supervised machine learning algorithm made up of decision trees. Random Forest is used for both classification and regression—for example, classifying whether an email is “spam” or “not spam”.\\
    \bottomrule
 \end{tabular}
\end{table*}

\subsubsection{Conventional Data Sources}
This section describes three non-deep learning methods proposed by researchers that focus on conventional data sources. Table 8 represents the proposed methods and their comparison with each other and ground truth, as well as future work and limitations.

\begin{table*}
\small
  \caption{Outbreak Detection Using Non-deep Learning Methods and Conventional Data Sources}
  \label{tab:commands}
   \begin{tabular} {p{0.2cm}| p{3cm} p{3cm} p{4cm} p{4cm}} 
    \toprule 
    {\bfseries No} & {\bfseries Models}& {\bfseries Data Source} & {\bfseries Result} & {\bfseries Limitations/Future work} \\
    \midrule

 \multirow{ 8}{*}{ 1}    
    &
    \begin{itemize}
        \item LASSO regression  
        \cite{chen_utility_2018} (2018)
    \end{itemize}
   & 
   National Institute of Infectious Diseases (NIID) - Japan,
    Bureau of Epidemiology, Ministry of Public Health - Thailand, Ministry of Health- Singapore, Taiwan National Infectious Disease Statistics System - Taiwan, Weather Underground - Taiwan, Thailand, Singapore, Japan Meteorological Agency - Japan
   &
    Regression models utilizing LASSO can notably enhance predictive accuracy for certain diseases using a specific set of variables; however, for other diseases, simpler models yield comparable outcomes to more intricate ones. Regular implementation of models based on this approach reveals superior short-term disease predictions compared to long-term forecasts, implying the necessity for rapid response capabilities in public health agencies to address early indications of potential infectious disease outbreaks.
    &
    limitations include its divergence from traditional epidemic models like compartmental and network models, as well as its reliance on capital city weather data, overlooking regional climatic variations; future accuracy may increase with finer data resolution.\\
    \midrule
    \multirow{ 8}{*}{ 2}  
& \begin{itemize}
 \item Gaussian process regression+LASSO  
 \item Linear regression
 \item ANN
 \item SVR 
 \item SARIMA
 \cite{chen2019predicting} (2019)
 \end{itemize}
    &
    Influenza-like illness (ILI) Shenzhen (CDC)
    &
    Gaussian process regression + LASSO model outperforms other models in terms of one-week-ahead prediction of influenza-like illness.
    &  
    As a future work, the authors intend to examine the efficacy of the proposed method in additional cities within the next few years. In addition, spatial information will be incorporated to model the effect of meteorology on influenza's spatial-temporal spread. \\
 \midrule
    \multirow{ 7}{*}{ 3}  
    & \begin{itemize}
 \item ARIMA
 \item XGBoost
 \cite{fang_application_2022} (2022)
 \end{itemize}
 & 
 USA COVID-19 cases
 and vaccination From CDC
 & 
 Based on the daily case numbers from the previous 7 days, the fit and prediction accuracy for the following 14 days are significantly improved by the XGBoost model.
 &
 This work has the following limitations: first, the study period was relatively short and should have been expanded to better reflect the future development of COVID-19 in the United States. And secondly, the XGBoost model was created using prevaccination-induced herd immunity as its foundation. Consequently, as the incidence of transmissible variants rises, the accuracy of predictions may diminish.\\
\bottomrule
 \end{tabular}
\end{table*}

In the initial analysis, Chen et al. (2018) \cite{chen_utility_2018} delved into the efficacy of employing the machine learning LASSO method for predictive modeling of various pathogens across diverse climatic conditions. Their approach involved creating distinct LASSO models for each disease, country, and forecast window, thereby assessing the relative significance of predictors under varying contextual settings. The results illuminated distinct cyclical patterns in climatic variables and disease incidence, particularly evident in regions geographically distant from the equator. Furthermore, the study underscored the inherent challenges of achieving accurate long-term predictions while highlighting the superior performance of short-term projections, underscoring the crucial need for swift responses to early warning signals.

Turning attention to the subsequent investigation, the study by Chen et al. (2019) \cite{chen2019predicting} centered on influenza forecasting utilizing Gaussian process regression techniques. Their innovative approach incorporated meteorological factors to gauge their influence on influenza transmission dynamics. The integration of L1-regularization aided in identifying key explanatory variables contributing to the predictive model. Notably, adjustments were made to the time covariance function to account for non-stationarity and seasonal patterns. Through rigorous comparisons with conventional statistical models, the proposed model showcased its prowess in predicting influenza-like illness (ILI) one week ahead, providing valuable insights into imminent disease trends.

The final review article showcased the contributions of Fang et al. (2022) \cite{fang_application_2022}, who devised ARIMA and XGBoost models to predict COVID-19 trends in the USA. A comprehensive assessment of both models' fitting capabilities and prediction accuracies was conducted, resulting in a pivotal discovery. The XGBoost model emerged as a game-changer, significantly enhancing fit and prediction accuracy owing to its adeptness in capturing non-linearities within the temporal dynamics of COVID-19 cases.

\subsubsection{Social Media and Internet Based  Data Sources}

Public health authorities employ Epidemic Intelligence (EI) to acquire data on disease activity, early warning, and infectious disease outbreaks \cite{linge_internet_2009,kaiser_different_2006,paquet_epidemic_2006,coulombier_epidemiological_2002}. From a variety of formal and increasingly informal sources, EI systems routinely compile official reports and rumors of probable outbreaks \cite{noauthor_epidemic_nodate}. To detect information on disease outbreaks, tools like the Global Public Health Intelligence Network (GPHIN) and Medisys collect data from international media sources like news wires and websites \cite{linge_internet_2009}. 
Google's Flu Trends research, which evaluated flu activity by compiling real-time web search queries for flu-related terms \cite{philanthropy_programs_for_underserved_communities_-_googleorg_philanthropy_nodate}, has shown how these systems might be improved. The information kept in commercial search query logs, which may be linked to EI systems, has the issue of not being publicly accessible. However, the rise of user-generated content on social networking sites like Facebook and Twitter gives EI systems a very easy way to access data on current online activity. There are already more than 15 million unique users per month using Twitter \cite{bohlin_tracking_2012}, a micro-blogging site that enables users to publish and read 140-character messages, or "tweets," from other users \cite{twittercrunchbasecompanyprofilefundingtwitter_nodate}. Twitter gives third parties the ability to search user messages and retrieve the text along with user-specific data, such as the poster's location, in a format that is simple to store and analyze. Table 9 depicts the proposed methods and their comparison with others and ground truth, as well as future work and limitations.

\begin{table*}
\small
  \caption{Outbreak Detection Using Non-deep Learning Methods and Social Media or Internet Data Sources}
  \label{tab:commands}
   \begin{tabular} {p{0.15cm}| p{3cm} p{3cm} p{4cm} p{4cm}} 
    \toprule 
    {\bfseries No} & {\bfseries Models}& {\bfseries Data Source} & {\bfseries Result} & {\bfseries Limitations/Future work} \\
    \midrule

 \multirow{ 8}{*}{ 1}    
    &
    \begin{itemize}
        \item SVM 
        \item Naïve Bayes
        \item Random Forest 
        \item Decision Tree
        \cite{jain_effective_2015} (2015)
    \end{itemize}
   & 
   Keywords from tweets and Keywords from RSS feeds for H1N1 or Swine flu in India
   &
    The analysis of data sets and classification results in the development of an early warning system that detects an impending spike in an epidemic before the official surveillance systems are examined. SVM performed best in classification as well.
    &
    Not Available \\
    \midrule
    \multirow{ 8}{*}{ 2}  
& \begin{itemize}
 \item Decision Tree  
 \item Random Fores
 \item SVM
 \item Naïve Bayes 
 \item AdaBoost
 \item KNN
 \item FastText
 \cite{alessa_preliminary_2019} (2019)
 \end{itemize}
    &
    Twitter influenza surveillance dataset
    &
    In predicting flu outbreaks, the combination of FastText (FT) classification with a regression model outperforms other classification algorithms.
    &  
    Not Available\\
 \midrule
    \multirow{ 7}{*}{ 3}  
    & \begin{itemize}
        \item Decision trees
        \item Random forests
        \item SVM
        \item SVM-Perf
        \cite{joshi_automated_2020} (2020)
    \end{itemize}
 & 
 Twitter posts between 2011 and 2014 in key cities of the affected region in West Africa
 & 
The results show that the adapted architecture using SVM-Perf obtains more relevant alerts than the others.
 &
 The proposed architecture could be adapted for other social media platforms or disease types in the future.\\
 \midrule
    \multirow{ 4}{*}{ 4}  
    & \begin{itemize}
    \item SVM
    \item Naïve Bayes
    \cite{fakhry_tracking_2020} (2020)
    \end{itemize}
 & 
 Twitter using keywords, language, 
Geolocation related to the Covid-19
 & 
 The Naïve Bayes classifier performed better than SVM at identifying tweets related to Covid-19.

 &
Not Available\\
\midrule
    \multirow{ 7}{*}{ 5}  
    & \begin{itemize}
    \item RF
    \item KNN
    \item SVM
    \item DT
    \cite{amin_early_2021} (2021)
    \end{itemize}
 & 
 Tweets on dengue and flu from Twitter
 & 
 In terms of finding tweets related to seasonal outbreaks, the results showed that the RF classifier outperformed SVM, DT, and KNN.
 &
There are some drawbacks to the proposed model. Because supervised learning was used in this study, the data used for model training had to be labeled. To avoid the need for labeled data, the model should be trained unsupervised. In the future, the proposed model could be used as a surveillance system to detect the spread of coronavirus and COVID-19.\\

\bottomrule
 \end{tabular}
\end{table*}

The initial study under review, conducted by Jain and Kumar in 2015 \cite{jain_effective_2015}, devised a method for tracking the flu epidemic in India during February to March 2015, employing Twitter content. An innovative approach based on dynamic keywords sourced from RSS feeds was introduced to retrieve tweets via Twitter. Data spanning 60 days, commencing from February 1, 2015, and concluding on March 31, 2015, was collected from Twitter. This dataset was instrumental in monitoring key terms associated with "H1N1" or "swine flu" over time. The authors conducted content analysis of tweets and scrutinized states most significantly impacted by the rapidly spreading flu.

Employing sentiment analysis and a count-based technique, the dataset was dissected to unravel pivotal aspects during the Influenza-A (H1N1) pandemic. The study systematically explored various parameters to extract pertinent information about the disease and gauge the general awareness surrounding it. Classification was deployed to differentiate real-time data from noise or irrelevant tweets. Four distinct algorithms (SVM, Naïve Bayes, Random Forest, and Decision Tree) were employed for classification, with the SVM classifier demonstrating superior performance.

This investigation highlighted the potential of using social media to comprehend public health dynamics, illustrating Twitter's utility in detecting disease outbreaks by analyzing data generated within the realm of social media.

Shifting the focus to the work by Alessa and Faezipour in 2019 \cite{alessa_preliminary_2019}, the authors formulated a comprehensive framework for predicting outbreaks in 2019. Comprising three key modules—text classification, mapping, and linear regression—the framework was designed to forecast weekly flu rates. The text classification module harnessed sentiment analysis and predefined keyword occurrences. Twitter influenza surveillance datasets were gathered and various classifiers, including FastText (FT) and six conventional machine learning algorithms, were evaluated to ascertain optimal effectiveness for the framework. The classified Twitter documents and historical CDC data were then fed into a linear regression-based module for predicting weekly flu rates. Remarkably, the proposed FastText (FT) classification emerged as the most efficient and accurate model. Impressively, the final flu trend prediction based on Twitter documents exhibited a robust Pearson correlation of 96.29 percent with the CDC's actual data from the initial months of 2018.

Moving on to the research by Joshi et al. in 2020 \cite{joshi_automated_2020}, two variations of an existing surveillance architecture were examined. The first version aggregated tweets related to various symptoms, while the second version considered each symptom separately before amalgamating the set of alerts produced by the architecture. This study utilized a database of tweets from impacted areas in West Africa between 2011 and 2014, focusing on the Ebola symptoms of fever and rash. The results led to two crucial conclusions: social media provided an early warning three months prior to the 2014 Ebola pandemic, and data aggregation could potentially yield more frequent notifications than alert aggregation. The SVM and SVM-Perf classifiers were employed for personal health mention categorization. The findings indicated that the modified architecture utilizing SVM-Perf produced more relevant alerts compared to the SVM-only architecture.

In the study conducted by Fakhry and colleagues in 2020 \cite{fakhry_tracking_2020}, a dual machine learning approach was employed to analyze present and future Covid-19 case data using publicly available social media information. Through a combination of machine learning and classical data mining techniques, disease cases were estimated based on social media content in a specific geographic region. The sentiment analysis was leveraged to gauge the public's perception of disease awareness in the same location. Employing specific keywords, tweets related to Coronavirus were extracted from Twitter and classified using Naïve Bayes and SVM classifiers. Strong correlation was observed between classified tweets and real-world data.

Lastly, Amin et al.'s algorithm in 2021 \cite{amin_early_2021} aimed to detect seasonal outbreaks using Twitter data through machine learning approaches. Two categories of tweets—disease-positive and disease-negative—were employed to identify outbreaks related to dengue and flu. Machine learning algorithms including Random Forest (RF), K-Nearest Neighbor (KNN), Support Vector Machine (SVM), and Decision Tree (DT) were harnessed, along with Term Frequency and Inverse Document Frequency (TF-IDF) for feature extraction. Notably, the RF classifier outperformed others in terms of accuracy, precision, recall, and F1 measure. Despite its findings, the model employed supervised learning, highlighting a need for further exploration of unsupervised training approaches.

\subsection{Deep Learning Time Series Methods}

Deep learning has been applied in healthcare through medical imaging, chatbots for pattern detection in patient complaints, algorithms for cancer detection, and systems identifying rare diseases or pathology. It provides valuable insights to medical professionals, enabling early problem detection and more personalized care. Table 10 highlights common deep learning methods used for outbreak detection with conventional and internet time series data.

\begin{table*}
\small
  \caption{Deep Learning Time Series Methods Summary}
  \label{tab:commands}
  \begin{tabular}{p{0.25\linewidth}p{0.7\linewidth}}
    \toprule
    Model & Interpretation\\
    \midrule
    \texttt{LSTM \cite{gers_applying_2002,kara_multi-step_2021,lee_predicting_2021,absar_efficacy_2022,jia_predicting_2019}} & LSTM stands for Long short-term memory. LSTM cells are used in recurrent neural networks that learn to predict the future from sequences of variable lengths. The main idea behind LSTM cells is to learn the important parts of the sequence seen so far and forget the less important ones. An LSTM network is a recurrent neural network (RNN) that processes input data by looping over time steps and updating the network state. The network state contains information remembered over all previous time steps. LSTM network can be used to forecast subsequent values of a time series or sequence using previous time steps as input.\\
    \midrule
    \texttt{Bi-LSTM \cite{shastri_chapter_2022,shastri_time_2020,arora_prediction_2020}}& Bidirectional LSTM, or biLSTM, is a sequence processing model that consists of two LSTMs: one taking the input in a forward direction, and the other in a backward direction. In problems where all timesteps of the input sequence are available, Bidirectional LSTMs train two instead of one LSTMs on the input sequence.\\
    \midrule
    \texttt{TBAT \cite{arunkumar2021time,zelek_why_2020,folorunso_comparison_2021,papastefanopoulos_covid-19_2020}}& TBATS model is a forecasting model based on exponential smoothing. The name is an acronym for Trigonometric, Box-Cox transform, ARMA errors, Trend, and Seasonal components. The TBATS model’s main feature is its capability to deal with multiple seasonalities by modeling each seasonality with a trigonometric representation based on the Fourier series.\\
    \midrule
    \texttt{N-BEATS \cite{zelek_why_2020,papastefanopoulos_covid-19_2020,jin_data-driven_2022}}& N-BEATS is a deep neural structure featuring backward and forward residual links and a deep stack of fully connected layers. It takes an entire historical data window and generates multiple forecast time points simultaneously through extensive use of fully connected layers. The architecture comprises interconnected blocks in a residual manner: the initial block models past and future data, while subsequent blocks focus on residual errors from the previous reconstruction, updating forecasts accordingly. This residual-based design enables a deep stack of blocks without gradient vanishing concerns and offers benefits akin to boosting/ensembling techniques, where predictions from various blocks are combined, with each block capturing different aspects of the forecast.\\
    \midrule
     \texttt{FB Prophet \cite{ai_application_2022,zhao_prediction_2022,aditya_satrio_time_2021,dash_intelligent_2021,battineni_forecasting_2020}}& FB Prophet is a procedure for forecasting time series data based on an additive regression model where non-linear trends are fit with yearly, weekly, and daily seasonality, plus holiday effects. It works best with time series that have strong seasonal effects and several seasons of historical data. it is great for stationary data. Stationary data is time series data that follow similar behavior and have the same statistical properties throughout time.\\
     \midrule
     \texttt{DeepAR \cite{arunkumar2021time,zelek_why_2020,folorunso_comparison_2021,papastefanopoulos_covid-19_2020,shaik_deep_2020,balaban_growth_2020,dash_bifm_2021}}& DeepAR, a supervised time series forecasting algorithm, employs recurrent neural networks (RNN) to generate both point and probabilistic predictions. This method considers not only past values but also incorporates additional covariates like dynamic historical features, static attributes, and future events. However, traditional techniques often treat each time series independently, missing out on potential cross-learning opportunities and valuable information relevant to the specific use case.\\
     \midrule
     \texttt{NARNN \cite{zhou_time_2018,saliaj_artificial_2022,kirbas_comparative_2020}}& The Nonlinear Autoregression Neural Network (NARNN) model is a technique that performs nonlinear regression through the neural network.\\
    \bottomrule
 \end{tabular}
\end{table*}

\subsubsection{Conventional Data Sources}

Deep learning in healthcare has received a lot of attention in recent years as a crucial tool for assisting in clinical decision-making and disease diagnosis \cite{topol_high-performance_2019,tomasev_clinically_2019}. In late December 2019, a Canadian company (Blue Dot) correctly notified the location of Covid-19 outbreak, demonstrating the effectiveness of AI and deep learning in the current pandemic for outbreak prediction. The development of image verification to differentiate COVID-19 pneumonia from other benign respiratory illnesses also benefited from AI \cite{wang_deep_2020}. We provide a summary of eleven recent studies that used deep learning to anticipate and identify outbreaks in this section. The methods, data sources, and findings used by earlier researchers to show and compare the effectiveness of their methods are summarised in Table 11.
\begin{table*}
\tiny
  \caption{Outbreak Detection Using Deep Learning Methods and Conventional Data Sources}
  \label{tab:commands}
   \begin{tabular} {p{0.2cm}| p{4cm} p{2.5cm} p{2.5cm} p{5cm}} 
    \toprule 
    {\bfseries No} & {\bfseries Models}& {\bfseries Data Source} & {\bfseries Result} & {\bfseries Limitations/Future work} \\
    \midrule

 \multirow{ 8}{*}{ 1}    
    &
    \begin{itemize}
        \item ARIMA 
        \item Holt–Winters additive
        \item TBAT 
        \item Facebook’s Prophet
        \item DeepAR
        \item N-Beats
        \cite{papastefanopoulos_covid-19_2020} (2020)
    \end{itemize}
   & 
   Novel Corona Virus 2019 dataset And
   population-by-country dataset
   &
    The ARIMA, Holt–Winters, and TBAT outperform the deep learning models due to the small dataset size.
    &
    Future works include enhancing prediction accuracy, potential improvements involving creating model ensembles to minimize errors, adopting multivariate time series modeling considering various pandemic-related factors, and transferring knowledge across countries. \\
    \midrule
    \multirow{ 5}{*}{ 2}  
& \begin{itemize}
 \item  DeCoVNet
 \cite{zheng_deep_2020} (2020)
 \end{itemize}
    &
    Data from the National Health Commission of the People’s Republic of China Diagnosed with COVID-19
    &
    DeCoVNet shows promise for precise and swift COVID-19 diagnosis, aiding frontline medical staff and global epidemic control.
    &  
   Limitations include Network design and training that could improve, data from a single hospital lacks cross-validation, and the algorithm's COVID-19 diagnosis is hindered by deep learning's early-stage explainability.\\
 \midrule
    \multirow{ 7}{*}{ 3}  
    & \begin{itemize}
        \item ARIMA
        \item EHW
        \item Linear Regression
        \item SVM Regression
        \item FDTL
        \item M5P Regression
        \item PNN
        \item PNN+cf
        \cite{fong_finding_2020} (2020)
    \end{itemize}
 & 
 2019-nCov from Chinese health authorities
 & 
The findings demonstrate that PNN+cf is effective in producing reliable forecasts during the crucial period of disease outbreaks when the samples are scarce.
 &
 Future work should analyze variations in forecasting outcomes among algorithms, test more panel selection methods, explore prominent algorithms' designs, and expand the methodology.\\
 \midrule
    \multirow{ 4}{*}{ 4}  
    & \begin{itemize}
    \item Stacked LSTM
    \item Bi-LSTM
    \item Conv-LSTM
    \cite{arora_prediction_2020} (2020)
    \end{itemize}
 & 
 Covid-19 data from the Ministry of Health and Family Welfare (Government of India)
 & 
 Bi-LSTM is better suited for prediction purposes of 1 to 7 days in the future.
 &
Not Available\\
\midrule
    \multirow{ 7}{*}{ 5}  
    & \begin{itemize}
    \item Stacked LSTM
    \item Bi-LSTM
    \item Conv-LSTM
    \cite{shastri_time_2020} (2020)
    \end{itemize}
 & 
 Indian data is sourced from the Ministry of Health and Family Welfare, and US data from the Centers for Disease Control and Prevention.
 & 
 ConvLSTM model outperformed the other two models for predicting the next month's case numbers.
 &
Future research: Analyzing COVID-19's economic impact across sectors and devising recovery strategies; investigating aerosol transmission and global case prediction.\\
\midrule
    \multirow{ 7}{*}{ 6}  
    & \begin{itemize}
    \item ARIMA
    \item SVR
    \item LSTM
    \item Bi-LSTM
    \item GRU
    \cite{shahid_predictions_2020} (2020)
    \end{itemize}
 & 
 Covid-19 dataset for different countries.
 & 
Bi-LSTM outperforms others in predicting Covid-19. Data for each country includes given cases for 110 days and must be predicted for the next 48 days.
 &
Not Available\\
\midrule
    \multirow{ 7}{*}{ 7}  
    & \begin{itemize}
    \item ARIMA
    \item NARNN
    \item LSTM
    \cite{kirbas_comparative_2020} (2020)
    \end{itemize}
 & 
Covid-19 data from European Center for Disease Prevention and Control for 8 different European countries (Denmark, Belgium, Germany, France, United Kingdom, Finland, Switzerland, and Turkey) 
& 
Based on the results, it was determined that the LSTM approach is significantly more successful than ARIMA and NARNN to predict the number of cases for the next 7 days.
 &
Not Available\\
\midrule
    \multirow{ 7}{*}{ 8}  
    & \begin{itemize}
    \item Bi-LSTM
    \item ARIMA
    \item (SMA-6), 
    Double Exponential 
    \item D-EXP-MA
    \cite{said_predicting_2021} (2021)
    \end{itemize}
 & 
 Covid-19 case numbers for different countries from their health authorities' websites
 & 
Bi-LSTM outperforms other models using 6 previous days' cases to predict the next day's case.
 &
Not Available\\
\midrule
    \multirow{ 5}{*}{ 9}  
    & \begin{itemize}
    \item LSTM
    \item XGBoost
    \cite{luo_time_2021} (2021)
    \end{itemize}
 & 
 USA COVID-19 cases were collected from the World Health Organization website.
 & 
LSTM outperforms XGBoost in predicting the next day case number based on the previous 7 days of data.
 &
Future research: Modeling lockdown-easing scenarios and their impact on case trends; assessing vaccination effects on COVID-19 outbreak in Qatar.\\
\midrule
    \multirow{ 7}{*}{ 10}  
    & \begin{itemize}
    \item LSTM
    \item GRU
    \item CNN
    \cite{abbasimehr_novel_2022} (2022)
    \end{itemize}
 & 
COVID-19 datasets for 10 countries from Humanitarian Data Exchange (HDX).
 & 
The results showed that after data augmentation, the performance of the LSTM and CNN models significantly improved. Furthermore, the proposed method achieves an average performance for GRU.
 &
Limitations include lack of intervention/vaccination data and sole reliance on infection time series. Future: Investigate time series augmentation methods like dynamic time-warping barycentric averaging.\\
\midrule
    \multirow{ 7}{*}{ 11}  
    & \begin{itemize}
    \item LSTM
    \item CDC ensemble model
    \cite{du2022deep} (2022)
    \end{itemize}
 & 
CDC and CSSE Johns Hopkins University data related to infectious diseases in the United States
 & 
The study demonstrated that the LSTM model outperformed existing methods, including the CDC ensemble model, in forecasting COVID-19 cases and deaths, highlighting the effectiveness of incorporating diverse data sources and advanced deep learning techniques.
 &
One limitation of this study is the reliance on self-reported data, which may introduce bias\\
\bottomrule
 \end{tabular}
\end{table*}

The ongoing COVID-19 pandemic has triggered substantial societal disruptions, compelling governments to implement drastic measures to curb its spread. Anticipating the outbreak's peak could significantly mitigate its impact, enabling governments to tailor policies, plan preventive actions, enhance public health communication, and bolster healthcare systems. As outlined in Table 11, the initial review by Papastefanopoulos et al. in 2020 \cite{papastefanopoulos_covid-19_2020} examined the accuracy of diverse time series modeling techniques for identifying coronavirus epidemics across ten countries with the highest confirmed cases by May 4, 2020. Six time series methodologies—ARIMA, Holt-Winters additive model (HWAAS), TBAT, Facebook's Prophet, DeepAR, and N-Beats—were developed and compared for each country. Publicly available datasets on virus progression and population size were employed for model development and analysis, acquired from "Novel Corona Virus 2019 Dataset" \cite{kaggle_novel_nodate} and "population-by-country dataset" \cite{population_by_country_2020} on kaggle.com.

Results showcased that while no single approach was universally optimal, conventional statistical methods like ARIMA and TBAT outperformed deep learning counterparts such as DeepAR and N-BEATS, which aligned with expectations given limited data availability. Specifically, ARIMA and TBAT exhibited superior performance across seven of ten cases. Statistical analysis employing Friedman's test and Holm's post-hoc analysis validated the superiority of TBAT over Prophet, DeepAR (Gluonts), and N-BEATS.

The COVID-19 outbreak placed significant strain on Hubei province, leading to a substantial volume of chest CT scans for suspected patients. The shortage of medical professionals exacerbated the risk of missed diagnoses for minor lesions.

The subsequent review, conducted by Zheng et al. in 2020 \cite{zheng_deep_2020}, introduced a 3D deep convolutional neural network (DeCoVNet) for COVID-19 detection through 3D CT volumes. The model encompassed lung segmentation via a pre-trained UNet, followed by a 3D deep neural network to predict COVID-19 infection likelihood. A dataset of 499 CT volumes for training and 131 volumes for testing were used. The algorithm achieved high ROC AUC, PR AUC, accuracy, sensitivity, specificity, and predictive values for COVID-positive and COVID-negative classification.

Fong et al. in 2020 \cite{fong_finding_2020} addressed the challenge of forecasting at the initial stages of an epidemic with scarce data. Their approach, Group of Optimized and Multi-Source Selection (GROOMS), combined multiple forecasting models, including polynomial neural networks (PNN), for group predictions. Experiments showed that PNN, particularly PNN+cf, excelled in generating accurate forecasts even with limited data.

Arora et al. in 2020 \cite{arora_prediction_2020} employed deep learning models, specifically recurrent neural network (RNN)-based LSTM variations, to forecast COVID-19 cases in Indian states. Accurate short-term predictions were achieved using Bi-directional LSTM.

Shastri et al. \cite{shastri_time_2020} employed LSTM, Stacked LSTM, and Convolutional LSTM to comparatively analyze COVID-19 cases in India and the USA. Convolutional LSTM outperformed the other models, providing accurate predictions and highlighting the significance of various factors beyond model choice.

Shahid et al. in 2020 \cite{shahid_predictions_2020} presented forecast models for COVID-19 cases in ten countries, utilizing ARIMA, SVR, LSTM, Bi-LSTM, and GRU models. Bi-LSTM consistently demonstrated superior performance, offering robust and accurate predictions for pandemic planning.

Kirbas et al. in 2020 \cite{kirbas_comparative_2020} modeled COVID-19 cases using ARIMA, NARNN, and LSTM techniques. LSTM emerged as the most precise model for short-term prediction, providing insights into the outbreak's trajectory.

Said et al. in 2021 \cite{said_predicting_2021} introduced a deep learning approach for predicting daily cumulative COVID-19 cases, grouping countries with similar characteristics. Their method significantly improved prediction performance compared to existing techniques.

Luo et al. in 2021 \cite{luo_time_2021} harnessed LSTM and XGBoost algorithms for predicting daily confirmed cases in the US. Their analysis emphasized the importance of precautionary measures and accurate data for meaningful predictions.

Abbasimehr et al. in 2022 \cite{abbasimehr_novel_2022} employed time series augmentation techniques to enhance deep learning model accuracy. The proposed augmentation approach exhibited superior performance across multiple countries, improving forecasting accuracy.

Lastly, Du et al. in 2022 \cite{du2022deep} presented a novel multi-stage deep learning model for forecasting COVID-19 cases and deaths in the United States. The model, referred to as the LSTM model, utilizes a comprehensive dataset encompassing epidemiological, mobility, survey, climate, demographic, and genomic data. Through rigorous evaluation, the LSTM model consistently outperforms the CDC ensemble model for all evaluation metrics, particularly in longer-term forecasting. The study emphasizes the importance of considering various factors when forecasting COVID-19 outbreaks. These factors include outbreak phase, location, time, and the availability of genomic data. The authors highlight the need for careful model selection and evaluation to ensure accurate and reliable predictions. Additionally, the study demonstrates the value of incorporating genomic surveillance data to enhance forecasting accuracy, especially during periods of emerging variants.

\subsubsection{Social Media and Internet Based  Data Sources}

Social media allows users to share news, ideas, and opinions globally, and researchers can automate the collection and analysis of posts for more effective disease surveillance \cite{xie_detecting_2013}. Social media and internet search data have been successfully used to monitor outbreaks like Zika, Dengue, MERS, Ebola, and COVID-19 \cite{al2016potential}. This section reviews various outbreak detection methods, evaluation techniques, and challenges in using social media and internet-based data. It also presents findings on two deep learning methods, their data sources, results, and limitations, as shown in Table 12.

\begin{table*}
\small
  \caption{Outbreak Detection Using Deep Learning Methods and Social Media or Internet Data Sources}
  \label{tab:commands}
   \begin{tabular} {p{0.3cm}| p{2cm} p{3.3cm} p{3cm} p{3.5cm}} 
    \toprule 
    {\bfseries No} & {\bfseries Models}& {\bfseries Data Source} & {\bfseries Result} & {\bfseries Limitations/Future work} \\
    \midrule

 \multirow{ 8}{*}{ 1}    
    &
    \begin{itemize}
        \item DNN 
        \item LSTM
        \item ARIMA 
        \item OLS
        \cite{chae2018predicting} (2018)
    \end{itemize}
   & 
   KCDC, search query data from South Korean-specific search engines, Twitter social media big data, and weather data such as temperature and humidity
   &
    Deep learning models DNN and LSTM predicted infectious diseases with a 7-day lag significantly better than OLS and ARIMA models.
    &
    Limitations include three factors: the study's brief data collection period, regionally combined predictions, and consideration of a constrained set of deep learning model parameters.\\
    \midrule
    \multirow{ 8}{*}{ 2}  
    & \begin{itemize}
    \item SSL
    \item SVM
    \item DNN
    \cite{yang_modified_2020} (2020)
     \end{itemize}
    &
    Articles and reports provided by Medisys related to disease
    &
    SVM, which consistently performs well across fields; SSL, which performs well when label imbalanced data sets are used; and DNN, a trending method with outstanding performance, were used to predict disease occurrence.
    &  
    Limitations include requiring systematic data timeframe selection, especially for seasonal diseases. Initial data limitations and model performance suggest potential for improvement. Exploring similar disease patterns in other nations and integrating global air passenger data could enhance future predictions.\\
\bottomrule
 \end{tabular}
\end{table*}

The first review discusses the work by Chae et al. in 2018 \cite{chae2018predicting}, where the researchers aimed to predict infectious diseases using deep learning algorithms by incorporating various sources of data, including infectious disease occurrence data from the Korea Center for Disease Control (KCDC), search query data from South Korean-specific search engines, Twitter social media data, and weather data like temperature and humidity. The search queries used included both the disease's name and its symptoms. The study developed an infectious disease monitoring model by combining non-clinical search data, Twitter data, and weather data. The researchers created various Ordinary Least Squares (OLS) models with different combinations of variables and evaluated their explanatory power using corrected R-squared values. They introduced a lag parameter of 1-14 days for infectious diseases and selected a seven-day lag as the best parameter for prediction. They built OLS, ARIMA, Deep Neural Network (DNN), and Long Short-Term Memory (LSTM) models using the chosen parameters. Comparing these models, they found that DNN and LSTM models outperformed OLS and ARIMA models. DNN was more consistent when diseases were spreading, while LSTM was more accurate. The authors suggested that their models could help reduce reporting delays in existing surveillance systems, leading to cost savings.

The second review pertains to the research by Kim and Ahn in 2021 \cite{kim_infectious_2021}. They created three machine learning models to detect early outbreaks of infectious diseases using disease-related media articles and reports. The models used were Support Vector Machine (SVM), Semi-Supervised Learning (SSL), and Deep Neural Network (DNN). The authors collected Medisys data from January to December 2019, including article titles, descriptions, published dates and times, disease information, and geographic coordinates. The daily article counts related to each disease by country were structured as a numerical dataset. The models were trained on data from one three-month period and tested on the subsequent three-month period for prediction. SSL showed the best performance, followed by SVM and DNN. All three models achieved average accuracies greater than 0.7 and F1 scores greater than 0.75. The proposed models were seen as useful for preparing for future infectious disease outbreaks, particularly in countries with limited disease surveillance systems.

\subsection{Hybrid Time Series Methods}
Methods or approaches that are made up of two or three methods or algorithms are referred to as hybrid methods. This section elaborates on some of the strategies presented by authors for outbreak detection utilizing conventional and internet-based data. Table 13 depicts some of the hybrid models used by authors to detect outbreaks from conventional and social media/internet time series data.  

\begin{table*}
\small
  \caption{Hybrid Time Series Methods Summary}
  \label{tab:commands}
  \begin{tabular}{p{0.25\linewidth}p{0.7\linewidth}}
    \toprule
    Model & Interpretation\\
    \midrule
    \texttt{VAR-LSTM \cite{afzali_hybrid_2020}} & In this method the data are first trained by using the Vector Auto-regressive (VAR) technique, then the outputs are used as the inputs for the Long Short-Term Memory (LSTM) networks by using a deep learning (DL) approach.\\
    \midrule
    \texttt{CNN-LSTM \cite{ketu_india_2022,dastider_integrated_2021,zain_covid-19_2021,dairi_comparative_2021,biswas_lstm-cnn_2022,muhammad_cnn-lstm_2022}}& In this method LSTM models can easily capture sequence pattern information, but they are tailored to deal with temporal correlations and only use the features specified in the training set. Another popular deep learning method is convolutional neural networks (CNNs). CNN models are capable of filtering out noise in the input data and extracting more valuable knowledge for the final forecasting model.\\
    \midrule
    \texttt{SVM-RBM \cite{xiao2016detecting,sun_trends_2017}}& Restricted Boltzmann Machines are stochastic two-layered neural networks that belong to a category of energy-based models that can detect inherent patterns automatically in the data by reconstructing input. They have two layers visible and hidden. RBM used to serve as features auto encoding processor for SVM.\\
    \bottomrule
 \end{tabular}
\end{table*}

\subsubsection{Conventional Data Sources}
The five most recent and innovative hybrid methods for outbreak detection are the focus of this section. Table 14 includes a column for future work and limitations that helps researchers come up with some ideas for their work while also summarising these methods' names, data sources, and results.

\begin{table*}
\small
  \caption{Outbreak Detection Using Hybrid Methods and Surveillance Data}
  \label{tab:commands}
   \begin{tabular} {p{0.2cm}| p{4cm} p{2cm} p{4cm} p{4cm}} 
    \toprule 
    {\bfseries No} & {\bfseries Models}& {\bfseries Data Source} & {\bfseries Result} & {\bfseries Limitations/Future work} \\
    \midrule

 \multirow{ 8}{*}{ 1}    
    &
    \begin{itemize}
        \item RCAL-BiLSTM
        \item CNN, DTL, MLP, LR, XGBoost, KNN, DT
       , Xiaowei Xu et al. \cite{xu_deep_2020} models
        \cite{pustokhin_effective_2020} (2020)
    \end{itemize}
   & 
   Chest-X-Ray dataset
   &
    RCAL-BiLSTM model outperforms other models and can be incorporated in real-time hospitals to predict and classify the COVID-19 pandemic.
    &
    Not Available\\

    \midrule
    \multirow{ 5}{*}{ 2}  
    & \begin{itemize}
 \item SEIR
 \item LSTM
 \cite{yang_modified_2020} (2020)
 \end{itemize}
 & Domestic migration data and Covid-19 data of China
 & The dynamic SEIR model, combined with LSTM-based AI, effectively predicted COVID-19 epidemic peaks and sizes.
 &  Limitations include not considering variables like diagnostic capacity that could impact case numbers and not accounting for seasonal influences.
 \\
 \midrule
    \multirow{ 5}{*}{ 3}  
    & \begin{itemize}
    \item VAR-LSTM
    \item LSTM
    \cite{afzali_hybrid_2020} (2020)
     \end{itemize}
    &
    Johns Hopkins University and the Canadian Health Authority
    &
    The positive Covid-19 case numbers can be predicted by VAR-LSTM with high accuracy. It employed lag 2 data to forecast the case numbers for the upcoming 15 days, outperforming the LSTM in the process.
    &  
    Not Available\\
    \midrule
    \multirow{ 5}{*}{ 4}  
    & \begin{itemize}
    \item Original Ensemble forecasts, Rounded Ensemble forecasts
    \cite{ray2020ensemble} (2020)
     \end{itemize}
    &
    Center for Systems Science and Engineering (CSSE) at Johns Hopkins University
    &
    Accurate short-term predictions with well-calibrated intervals (92-96\% coverage), though accuracy declines for longer forecasts.
    &  
    Future work involves enhancing model calibration and expanding the ensemble to improve long-term forecast accuracy and robustness.\\
    \midrule
    \multirow{ 5}{*}{ 5}  
    & \begin{itemize}
    \item MLP
    \item ANFIS
    \item SIR/SEIR
    \cite{ardabili2020covid} (2020)
     \end{itemize}
    &
    Data collected from worldometers for five 
    countries, including Italy, Germany, Iran, USA, and China
    &
The research demonstrates the effectiveness of hybrid ML-statistical models in COVID-19 outbreak prediction.
    &  
    Future research should develop country-specific ML models to address the unique characteristics of COVID-19 outbreaks. Given the variability between outbreaks, creating a single global model with broad generalization may be challenging.\\
    \midrule
    \multirow{ 8}{*}{ 6}  
    & \begin{itemize}
    \item CNN-LSTM
    \item Other baselines (CNN, LSTM, ARIMA, FBProphet, LR, Ridge, Lasso, XGBoostR, AdaBoostR, RFR, GBR, ETR, BaggingR, GPR, SVR, DTR, KNNR)
    \cite{zain_covid-19_2021} (2021)
     \end{itemize}
    &
    WHO COVID-19 dashboard for all countries
    &
    When compared to 17 baseline time series forecasting models, the proposed CNN-LSTM model will outperform them all.
    &  
    Future work involves enhancing COVID-19 forecasting accuracy by incorporating additional data and external factors like seasonal changes, vaccination plans, and lockdowns. Exploring various resampling and restructuring methods is planned. Additionally, developing an uncertainty management strategy to quantify and convey pandemic information more effectively is essential.\\
    \midrule
    \multirow{ 8}{*}{ 7}  
    & \begin{itemize}
    \item LR
     ,Polynomial regression
    , LSTM
   , GRU
   , RNN
    , ARIMA
    , Prophet
    \item LSTM-GR
    \cite{sah_forecasting_2022} (2022)
     \end{itemize}
    &
    COVID-19 cases in India from Kaggle for each state
    &
    The study findings reveal that the proposed stacked LSTM-GRU model outperforms all other models.
    &  
    Future work could incorporate new components and algorithms into the hybrid model to address the prevalence of asymptomatic cases in India and improve result accuracy.\\
    \midrule
    \multirow{ 8}{*}{ 8}  
    & \begin{itemize}
    \item MLR
    , LSTM
    , Prophet
    , SEIR
    \item XGBoost-LSTM(XLM)
     \cite{guo_traffic_2022} (2022)
     \end{itemize}
    &
    Highly pathogenic infectious disease transmission dataset published by Baidu
    &
    The experiments show that the XLM prediction framework proposed in this paper outperforms other prediction methods.
    &  
    Future work includes enhancing the framework using big traffic data's distinct features to improve disease prediction accuracy, emphasizing key transmission attributes and expanding its applicability to support disease prevention.\\
    
\bottomrule
 \end{tabular}
\end{table*}

The first review discusses the work by Pustokhin et al. in 2020 \cite{pustokhin_effective_2020}, where they introduced the RCAL-BiLSTM model for COVID-19 diagnosis. The model involves preprocessing images using bilateral filtering (BF) to remove noise, followed by feature extraction using RCAL-BiLSTM, and classification using a softmax (SM) layer. The RCAL-BiLSTM model includes ResNet-based feature extraction, Class Attention Layer (CAL), and Bidirectional LSTM modules. The model's performance was evaluated on a Chest X-ray dataset, and it outperformed other models, achieving a higher F-score value of 93.10

The second review by Yang et al. in 2020 \cite{yang_modified_2020} integrated population migration data with a susceptible-exposed-infected-removed (SEIR) model and used an AI-based LSTM approach, trained on 2003 SARS data, to predict the progression of COVID-19. Their dynamic SEIR model effectively captured the epidemic peaks and trends, highlighting the crucial role of control measures implemented on January 23, 2020, in reducing the eventual size of the epidemic.

The third review pertains to research by Afzali et al. in 2020 \cite{afzali_hybrid_2020}, where they introduced a hybrid VAR-LSTM model for COVID-19 data modeling. The model combined Vector Auto-regressive (VAR) and Long Short-Term Memory (LSTM) techniques. The VAR approach was used to train the LSTM network, improving its ability to forecast confirmed COVID-19 cases. The hybrid VAR-LSTM model outperformed LSTM in terms of accuracy and performance when compared with actual case data.

In the fourth work by Evan L. Ray et al. in 2020 \cite{ray2020ensemble}, they analyzed the real-time application of an open, collaborative ensemble forecasting model to predict COVID-19-related deaths in the United States. Their study, spanning from April to July 2020, aggregated probabilistic forecasts from multiple models to create a weekly ensemble forecast. Each participating model provided predictions for one to four weeks ahead, including median estimates and a range of prediction intervals to account for uncertainty. The ensemble was constructed by averaging these predictions, providing a more robust and reliable forecast. The results indicated that these ensemble forecasts provided accurate short-term predictions, particularly within a one-week horizon, though the accuracy diminished at longer horizons. The ensemble's prediction intervals were well-calibrated, with observed outcomes falling within the predicted ranges, demonstrating the model's reliability for public health decision-making during the pandemic.

The fifth review focuses on the research proposed by Ardabili et al. in 2020 \cite{ardabili2020covid} that investigated the potential of combining machine learning (ML) with statistical models for predicting COVID-19 outbreaks. It compares ML algorithms like Multi-Layer Perceptron (MLP) with 8, 12, and 16 neurons, and Adaptive Network-Based Fuzzy Inference System (ANFIS) with Triangular, Trapezoidal, and Gaussian membership functions, with traditional epidemiological models, finding that ML models offer superior accuracy and generalization. The Grey Wolf Optimizer (GWO) is identified as the best parameter tuning method for statistical models. The logistic model consistently outperforms other statistical models. Both weekly and daily data sampling are effective for ML modeling. The study suggests ML can address challenges in traditional models, such as missing data. Integrating ML with SIR/SEIR models is proposed for enhanced accuracy and predictive power. Overall, the research demonstrates the effectiveness of hybrid ML-statistical models in COVID-19 outbreak prediction, offering a promising alternative to traditional approaches.

The sixth work was done on a time-series dataset by Zain et al. in 2021 \cite{zain_covid-19_2021}. The researchers developed a hybrid CNN-LSTM model to predict the number of confirmed COVID-19 cases. They used WHO COVID-19 dashboard data and compared their model with 17 baseline models. The CNN-LSTM model showed superior performance, indicating that combining both models in an encoder-decoder structure significantly improved forecasting accuracy. The CNN-LSTM model effectively handled noise through convolutional layers and captured short- and long-term relationships in the time series.

The seventh review discusses the work by Sah et al. in 2022 \cite{sah_forecasting_2022}, where they proposed a hybrid stacked LSTM-GRU model to predict COVID-19 cases in India. The stacked LSTM output was used as input for the GRU model, resulting in better prediction accuracy compared to other state-of-the-art models.

The last review discusses the work by Guo et al. in 2022 \cite{guo_traffic_2022}, where a mixed XGBoost-LSTM (XLM) framework was introduced to predict the spread of infectious diseases across multiple cities and regions. The framework utilized Baidu's infectious disease transmission dataset and combined K-means clustering, XGBoost modeling, and LSTM modeling. The XLM framework demonstrated better predictive performance compared to other methods for predicting infectious disease spread.

These reviews highlight various innovative approaches to utilizing deep learning models for COVID-19 diagnosis, case prediction, and transmission forecasting, each contributing valuable insights to the field of disease prediction and control.

\subsubsection{Social Media and Internet Based  Data Sources}
Health authorities maintain a surveillance system to limit the spread of infectious diseases, but missing and delayed information makes timely action challenging. Predicting illness patterns is also difficult due to their unpredictability. To address these issues, a data-driven infectious disease prediction model is needed, which could help reduce societal costs by forecasting trends. Recognizing this, researchers are increasingly focusing on data-driven studies to enhance current systems and develop new models \cite{balcan2009multiscale,colizza2007modeling,balcan2009seasonal,eubank2004modelling,ferguson2006strategies,epstein2007controlling,ciofi2008mitigation}. Many of these studies use large datasets like Internet search queries \cite{lampos2015advances,zhang2017monitoring,rohart2016disease,cho2013correlation,teng2017dynamic,dugas2013influenza}, which can be processed in near-real-time. Social media big data is also being considered. Table 15 presents recent studies utilizing hybrid models with social media data for outbreak detection.

\begin{table*}
\small
  \caption{Outbreak Detection Using Hybrid Methods and Social Media or Internet Data Sources}
  \label{tab:commands}
   \begin{tabular} {p{0.3cm}| p{3cm} p{3cm} p{3cm} p{3cm}} 
    \toprule 
    {\bfseries No} & {\bfseries Models}& {\bfseries Data Source} & {\bfseries Result} & {\bfseries Limitations/Future work} \\
    \midrule

 \multirow{ 8}{*}{ 1}    
    &
    \begin{itemize}
        \item Bayesian network (BN) \cite{xiao2016detecting} (2016)
        \item Hidden Markov Model(HMM)
        \item SVM 
        \item SVM-RBM
    \end{itemize}
   & 
   Influenza-related posts from the Sina microblog
   &
    Compared to other unsupervised and supervised models, the hybrid SVM-RBM model demonstrated greater robustness and efficacy.
    &
    The first step in future research is to determine whether the likelihood of being in an epidemic phase may be influenced by the rate from the previous week. The addition of a multivariate or spatial component to the proposed model would allow us to investigate any geographical disaggregation of the rates.\\
    \midrule
    \multirow{ 5}{*}{ 2}  
    & \begin{itemize}
    \item SVM
    \item Naïve Bayes
    \item RNN-LSTM \cite{adhikari_epidemic_2018}(2018)
     \end{itemize}
    &
    Twitter to extract the tweets with the epidemic name
    &
    Naïve Bayes with TF-IDF performed better than other methods and produced a superior outcome.
    &  
    Not Available\\
     \midrule
    \multirow{ 8}{*}{3}  
    & \begin{itemize}
    \item KNN
    \item CUSUM
    \cite{yeng2019k}(2019)
     \end{itemize}
    &
     Synthetic datasets simulating health status monitoring of 297 individuals with diabetes over 12 months, including attributes like infection status, location coordinates, date/time stamps, and personal features. Additionally, the study considered mobile application-generated data to capture blood glucose dynamics, contributing to the disease surveillance framework. 
    &
   The K-CUSUM hybrid framework achieved 99.52\% accuracy in classifying infections using KNN and accurately detected outbreak spikes through CUSUM algorithm, demonstrating effective cluster detection potential for disease surveillance.
    &  
    For limitation can mention challenges in achieving effective anonymization of geolocation data while preserving its utility for disease surveillance. 
    
    Future work may involve exploring unsupervised learning methods and assessing the system with empirical data for real-world implementation.\\
    \midrule
    \multirow{ 8}{*}{4}  
    & \begin{itemize}
    \item ODANN
    \item SVM, RF, ARIMA, 
 AutoARIMA, LSTM, Prophet
    \cite{chew2021hybrid}(2021)
     \end{itemize}
    &
     Twitter
    &
  ODANN demonstrates superior performance in predicting the global growth rate of COVID-19 cases compared to traditional time-series, deeply leaning and non-deep learning models.
    &  
  For Future work can mention the additional data sources: Incorporating other relevant data sources could further improve ODANN's performance.
Also the real-time predictions: Refining the model's ability to handle real-time data is crucial for timely decision-making.\\
\bottomrule
 \end{tabular}
\end{table*}

The first reviewed article by Xiao et al. \cite{xiao2016detecting} in 2016 demonstrates how social media data can be harnessed using data mining techniques to detect real-world phenomena. They treat microblog users as "sensors," using flu-related group posts as early indicators. By collecting microblog posts from Sina Weibo, they create a crowdsourced data approach. They propose supervised and unsupervised methods to estimate flu-infected individuals, focusing on sentiment scores in an unsupervised model. They enhance a supervised model with Conditional Random Fields (CRFs) for quicker flu outbreak detection. Their hybrid classification model using a Support Vector Machine and  Restricted Boltzmann Machine (SVM-RBM) classifies flu-related posts, enabling real-time flu detection. Both models outperform others when considering sentiment analysis and additional features. The proposed method is effective, as demonstrated by testing on real-time datasets for various applications.

The second article is a work that focuses on using the power of Text Analysis and Machine learning. Adhikari et al. \cite{adhikari_epidemic_2018} presented 2018 a noble method for predicting disease-prone areas. The primary objective of this work is to develop an Epidemic Search model utilizing the power of social network data analysis and then employing this data to provide a probability score of the spread and to analyze the areas likely to be affected by any epidemic spread.
The authors extracted tweets containing the name of the epidemic using Twitter. They have attempted to analyze and demonstrate how the model with various preprocessing techniques and algorithms predicts the output. Combining words-ngrammes, word embeddings, and TF-IDF with various data mining and deep learning algorithms such as SVM, Nave Bayes, and RNN-LSTM was utilized. As a result, Nave Bayes with TF-IDF performed superiorly compared to other algorithms.

The third review is related to the work conducted by Yeng et al. \cite{yeng2019k} in 2019. This study focused on the EDMON (Electronic Disease Monitoring Network) project, aiming to detect the early spread of contagious diseases. By employing a hybrid approach that combined K-nearness Neighbor (KNN) and cumulative sum (CUSUM) algorithms, the research achieved impressive results. The KNN algorithm demonstrated a remarkable 99.52\% accuracy in classifying infected individuals, while the CUSUM algorithm effectively identified outbreak clusters. This combination of spatial and temporal analysis, supported by machine learning techniques and data visualization tools, showcases the potential for accurate and timely disease outbreak detection. The study's approach holds promise for proactive public health intervention and safeguarding community well-being.

The last review focuses on a work done by Chew et al. in 2021 \cite{chew2021hybrid}, this research proposed a novel hybrid deep learning model, Optimized Data Assimilated Neural Network (ODANN), to predict the global growth rate of COVID-19 cases. By combining natural language processing (NLP) features extracted from Twitter data with data assimilation techniques, ODANN outperforms traditional time-series models, including ARIMA, RF, SVM, LSTM, AutoARIMA, and Prophet. ODANN achieves a Root Mean Squared Error (RMSE) of 0.00282 and a Mean Absolute Error (MAE) of 0.00214, demonstrating superior performance in predicting the global growth rate of COVID-19 cases. The study highlights the importance of considering social factors, such as public sentiment, in modeling infectious disease outbreaks.

\section{Challenges and Future Directions}\label{sec:challenge}

This article provides a comprehensive survey of disease outbreak detection methods that utilize diverse surveillance system data sources—a topic of significant importance. While we have previously presented tables summarizing limitations highlighted in prior research endeavors, here we will delve into several pivotal limitations in a broader context.

\begin{itemize}

 \item  \textbf {Background behavior and labeled data challenges:} 
The identification of data anomalies requires the establishment of a baseline for typical behavior. What constitutes a target signal in one context, such as a seasonal influenza epidemic, can become part of the background noise in another, complicating the dual purpose of biosurveillance systems for both detecting natural outbreaks and bioterrorism-related or pandemic diseases. Ensuring optimal sensitivity during and post a seasonal influenza outbreak necessitates purging bioterrorism monitoring of all seasonal influences. However, precisely timing naturally occurring disease outbreaks in specific areas remains an intricate task. The evaluation and comparison of detection algorithms suffer from a shortage of labeling. While epidemiologists can make informed estimates regarding the sequence in which people seek care, precisely measuring the time for infections to manifest specific behaviors and the subsequent interplay of these behaviors remains intricate. The highly fluid population under observation further complicates matters. The nonstationary, ever-evolving background behavior, influenced by population shifts, data reporting, hospital policies, and other factors, defies conventional modeling methods, perpetuating a dearth of high-quality training data.

\item  \textbf {Big data collection challenges and methods:}
Devising algorithms capable of dissecting vast, ever-evolving, and unstructured data prevalent in social media and online content presents ongoing challenges. Algorithms designed for real-time, accurate pandemic tracking must deliver high precision with minimal time delay. Ensuring data privacy and accessibility poses additional complexities. Social media APIs grant access solely to public data, limiting the data's scope. However, making formerly private data public can attract undesirable attention and harm(such as spamming and damage to one's reputation) \cite{hogben2007security,salathe_digital_2018}. Addressing these challenges requires expertise in epidemiology, analysis, and advanced computational skills. Conquering social media's constraints demands a meticulous methodology proficient in API system usage, managing substantial data streams, handling noisy data, and mitigating bias in data collection. Furthermore, refining disease outbreak detection accuracy hinges on the ability to pinpoint the timing and locations of message transmission. Geo-temporal disambiguation proves challenging, and text descriptions often engender significant ambiguity in place names.

 \item  \textbf {Data accuracy and relevancy challenges:}
Data and analysis quality exert a profound impact on the effectiveness of internet-based surveillance systems. Improving data quality and accuracy involves exploring methods to ascertain a user's geographical location based on profile information and language usage in texts. Addressing sample size limitations is also critical. Internet-based surveillance systems leverage rudimentary algorithms to identify infectious disease indicators. However, as social network usage surges and data grows exponentially, sophisticated algorithms are imperative for real-time pandemic extraction, analysis, detection, and tracking with unparalleled accuracy. Analyzing live, massive, and unstructured data mandates advanced computational linguistics. The systematic, generic approach of internet-driven surveillance renders it adaptable for monitoring a spectrum of infectious diseases. There's substantial headroom for progress in areas like detection (via high-filter methods), testing (as exemplified by Zika), and integration with traditional surveillance systems. To comprehensively fathom pandemic dynamics, flexible internet-based surveillance frameworks must unify data from diverse online platforms\cite{lombardo2009public,milinovich2014internet}. 

 \item  \textbf {Internet Search Context Interpretation:}
Interpreting the search context of specific queries or documents in internet-based data systems is challenging due to varied user motives—queries may pertain to drugs, symptoms, or illnesses for diverse reasons. The multifaceted meanings of a single word depending on its contextual use in surrounding text introduce complexity. Furthermore, a single disease may be referred to by multiple names and exhibit a wide range of symptoms. Future research should strive to minimize false alarms and the detection of insignificant events by epidemic systems.

\item  \textbf {Incorporating Spatial Information and Cluster Detection using Spatiotemporal Analysis:}
Amidst the myriad challenges of disease outbreak detection, an emerging avenue of exploration lies in the integration of spatial information and cluster detection through spatiotemporal analysis. Harnessing the geographical context of outbreaks can offer invaluable insights into disease spread patterns, aiding in the identification of localized clusters and hotspots. Such spatially aware methodologies provide a holistic view of epidemic dynamics, accounting for the interplay between geographical proximity, population density, and disease transmission.

Cluster detection techniques, particularly those rooted in spatiotemporal analysis, hold promise in unveiling hidden patterns within outbreak data. By identifying clusters of cases occurring within specific regions and timeframes, these methods offer a proactive approach to pinpointing disease emergence and spread. Spatiotemporal cluster detection not only aids in early warning systems but also facilitates targeted intervention strategies, optimizing resource allocation and containment efforts.

As the field advances, combining spatial information with machine learning algorithms becomes increasingly pertinent. Leveraging advanced computational techniques, such as geospatial machine learning, holds the potential to unravel complex relationships between disease dynamics, environmental factors, and human behavior. By developing models that fuse temporal trends, geographical proximity, and sociodemographic variables, researchers can attain a nuanced understanding of outbreak propagation.

\item  \textbf {Challenges and Future Directions in Spatial-Based Detection:}
However, incorporating spatial information introduces its own set of challenges. Spatial data often exhibits heterogeneity and uneven distribution, necessitating careful preprocessing and normalization. Additionally, reconciling disparate data sources, such as healthcare facility records, geolocation data, and social media feeds, requires sophisticated data fusion techniques. Effective utilization of spatial information also calls for domain expertise in geospatial analysis and disease epidemiology. Future research endeavors can focus on refining the accuracy and reliability of spatiotemporal cluster detection methods. This involves developing algorithms that account for varying spatial scales, temporal resolutions, and the incorporation of uncertainty measures. By integrating real-time data streams from remote sensing platforms, mobile devices, and wearable technology, researchers can enhance the timeliness and granularity of outbreak alerts.

Furthermore, ethical considerations must accompany the utilization of spatial data, especially when dealing with sensitive information about individuals' locations and movements. Striking a balance between public health interests and individual privacy remains a critical aspect of spatial-based surveillance.

\item  \textbf {Emerging Challenges and Areas of Exploration:}
While our survey comprehensively covers existing challenges, it's important to acknowledge that the field is ever-evolving. Emerging challenges may include issues related to cross-platform data integration, the ethical use of user-generated content, and the continuous refinement of algorithmic accuracy. Moreover, considering the rapid pace of technological advancements, new opportunities for harnessing novel data sources and employing advanced analytical techniques may arise, reshaping the landscape of disease outbreak detection.

By illuminating these intricate challenges, this survey equips researchers, practitioners, and policymakers with valuable insights as they navigate the dynamic realm of disease outbreak detection. As this field continues to evolve, this survey's relevance and impact are poised to endure.

\item  \textbf {Embracing Diverse Data Sources:} In light of the evolving landscape of data-driven research for outbreak detection, it is worth considering an expanded exploration of data sources beyond the dichotomy of social media/internet versus conventional data sources. As the field progresses, integrating emerging data streams such as satellite imagery, healthcare claims, retail purchases, environmental sensors, and genomic sequencing could offer a more comprehensive and dynamic approach to early outbreak detection. Additionally, datasets from platforms like YouTube, LinkedIn, and facebook, should be explored, especially considering the increasing limitations on Twitter API accessibility. These alternative social media data sources could provide valuable insights, further enhancing the diversity of inputs available for predictive modeling. This inclusive perspective acknowledges the multifaceted nature of outbreaks and leverages a diverse array of data sources to enhance accuracy, timeliness, and reliability in predicting and responding to health threats. Furthermore, ensuring the privacy and ethical considerations surrounding the collection and utilization of all these data sources remains a critical challenge that requires careful attention and robust solutions.

\end{itemize}
\section{Conclusion}\label{sec:conc}

This article presents a comprehensive overview of conventional and internet-based surveillance systems for identifying disease outbreaks and forecasting, offering an invaluable resource for researchers and practitioners alike. Our analysis covers a wide spectrum of literature, focusing on the key dimensions of data sources, methodologies, algorithmic comparisons, challenges, and future directions. By scrutinizing historical trends in the field, we aim to capture the evolution of outbreak detection techniques and forecasting, with a particular focus on data-driven approaches. Our review is structured around two pivotal axes: (1) data sources, and (2) statistical and machine learning techniques. A key finding of this survey is that 62\% of the studies utilized conventional data sources, while 38\% relied on social media or internet-derived data. Additionally, 20\% of the research employed statistical models, 28\% leveraged machine learning models (including both deep learning and non-deep learning techniques), and 16\% applied hybrid models. These insights underline the growing adoption of machine learning methods, alongside the continued relevance of traditional statistical approaches, in outbreak prediction and forecasting. This survey not only highlights the ongoing exploration of diverse frameworks and methodologies but also brings attention to the underutilized potential of hybrid and emerging models. Several critical factors influencing the performance of outbreak detection were identified, including data origin, dataset size, real-time data accessibility, machine learning hyperparameters, geolocation effects, and healthcare facility variables. Addressing these determinants, alongside the limitations and challenges outlined in the existing literature, will be crucial for future research. Our survey aims to close existing gaps by providing a cohesive synthesis of these methods and by suggesting potential directions for innovation.

The findings and discussions in this article will aid academics, researchers, clinicians, healthcare practitioners, and policymakers in navigating the evolving landscape of public health applications. As the domain of disease outbreak detection continues to expand, we hope this up-to-date survey fosters the exchange of ideas, encourages the development of novel techniques, and serves as a vital resource for shaping future research.

\bibliographystyle{unsrt}
\bibliography{sample-manuscript}

\end{document}